\documentclass[%
 reprint,
 amsmath,amssymb,
 aps,
]{revtex4-2}

\usepackage{graphicx}
\usepackage{dcolumn}
\usepackage{bm}
\usepackage{xcolor}
\usepackage{multirow}

\usepackage{hyperref}

\usepackage{mathtools}
\usepackage{empheq}

\usepackage{subcaption}
\newcommand{\bx}{\mathbf{x}}
\newcommand{\by}{\mathbf{y}}
\newcommand{\bz}{\mathbf{z}}
\newcommand{\bv}{\mathbf{v}}
\newcommand{\bk}{\mathbf{k}}
\newcommand{\bU}{\mathbf{U}}
\newcommand{\bZ}{\mathbf{Z}}
\newcommand{\cM}{\mathcal{M}}
\newcommand{\cQ}{\mathcal{Q}}
\newcommand{\R}{\mathbb{R}}

\newcommand{\ep}{\epsilon}
\newcommand{\abbr}[1]{{\tt{\uppercase{\scalebox{0.75}{#1}}}}}
\newcommand{\cN}{\mathcal{N}}
\newcommand{\red}[1]{{\color{red} #1}}

\begin{document}

\preprint{APS/123-QED}

\title{Turbulence enhancement of coagulation: \\ 
the role of eddy diffusion in velocity}
\thanks{Funded by the European Union (ERC, NoisyFluid, n. 101053472). Views and
opinions expressed are however those of the authors only and do not
necessarily reflect those of the European Union or the European Research
Council. Neither the European Union nor the granting authority can be held
responsible for them.}%

\author{Franco Flandoli}
\email{franco.flandoli@sns.it}
 \affiliation{Scuola Normale Superiore, Piazza Dei Cavalieri 7, Pisa PI 56126, Italy}

\author{Ruojun Huang}%
 \email{ruojun.huang@uni-muenster.de}
  \affiliation{Fachbereich Mathematik und Informatik, Universit\"at Münster, Germany}
 
 \author{Andrea Papini}
\email{andrea.papini@sns.it}
 \affiliation{Scuola Normale Superiore, Piazza Dei Cavalieri 7, Pisa PI 56126, Italy}



\date{\today}

\begin{abstract}
A Smoluchowski type model of coagulation in a turbulent fluid is given,
first expressed by means of a stochastic model, then in a suitable scaling
limit as a deterministic model with enhanced diffusion in the velocity
component. A precise link between mean intensity of the turbulent velocity
field and coagulation enhancement is obtained by numerical simulations, and a formula for the mean velocity difference, in agreement with the gas-kinetic model, is proved by a new method.
\end{abstract}

\maketitle


\section{Introduction}

Turbulence increases the relative velocity of particles suspended into a
fluid, favours their collision and thus increases the collision rate. A key
factor of the collision rate is the average relative velocity between
particles of mass $m_{1}$ and $m_{2}$:%
\begin{align}\label{vel-diff}
R_{m_{1},m_{2}}=\left\langle \left\vert \mathbf{v}_{1}-\mathbf{v}%
_{2}\right\vert \right\rangle .
\end{align}
This quantity is of major importance since it relates the properties of
particles and fluid to the intensity of the aggregation and thus it has been
extensively investigated in several works, based on various arguments and
models of turbulence, see for instance \cite{abrahamson1975collision, Ayala, Chun, Devenish, Falkovich, FalkPumir, Grabowski, Mehlig, Papini, 
Pumir,Reade, Saffman, Shima, Sundaram, Wang, wilkinson2006caustic,  Yeung}.
We shall add more specific comments below on some of these results in
connection with our own.

We propose a new modeling approach here. Many ingredients are classical, like
the fact that we use an \textit{inertial model} for particle motion (instead of a
model when particles are transported) where each particle moves following
Stokes' law%
\begin{equation}
\frac{d\mathbf{x}}{dt}=\mathbf{v},\qquad\frac{d\mathbf{v}}{dt}=\gamma\left(
\mathbf{U}\left(  t,\mathbf{x}\right)  -\mathbf{v}\right)
\label{particle-fluid}%
\end{equation}
(here $\gamma$ is the the damping coefficient and $\mathbf{U}\left(
t,\mathbf{x}\right)  $ is the fluid velocity), and \textit{Smoluchowski
equations} with a kernel depending on the relative velocity $\left\vert
\mathbf{v}-\mathbf{v}^{\prime}\right\vert $ to describe macroscopically the
system. The novelty is that we introduce a Boussinesq hypothesis, namely the
fact that a small-scale turbulence acts on particles as a dissipation. And the
key feature is that it acts as a \textit{dissipation in the velocity
component}, namely it spreads the distribution of particles in velocity (not
or not only in space). This spread increases the value of $R_{m_{1},m_{2}}$
and thus the collision rate.

In order to describe the equations we use and the results, let us recall a few
quantities associated to the particles and to the fluid. The damping
coefficient $\gamma$ appearing in equation (\ref{particle-fluid}) is given by
Stokes' law $\frac{6\pi r\mu}{m}$ where $r,m$ are the particle radius and mass
and $\mu$ is the dynamic viscosity of the fluid. If we denote by $\tau_{P}$
and $\tau_{\mathbf{U}}$ the relaxation times of the particle and of the fluid
respectively, we have $\gamma=\tau_{P}^{-1}$ and we define the Stokes number
as $St=\tau_{P}/\tau_{\mathbf{U}}=1/\left(  \gamma\tau_{\mathbf{U}}\right)  $.
When we want to stress the dependence of the damping coefficient $\gamma$ from
the mass $m$, we write $\gamma_{m}$; and similarly for $St_{m}$. Two relevant
quantities of the fluid for our study are the turbulence kinetic energy
$k_{T}=\frac{1}{2}\left\vert \overline{\mathbf{U}}\right\vert ^{2}$ and the
turbulent viscosity $\nu_{T}=\tau_{\mathbf{U}}k_{T}$. Our model is based on
the idealization that the turbulent small-scale fluid is white noise in time,
space-homogeneous, with intensity $\sigma$ (precisely, as a vector field, its
space-covariance matrix $C\left(  \mathbf{x}\right)  $ is assumed to have the
auto-covariance $C\left(  \mathbf{0}\right)  $ equal to $\sigma^{2}I_{d}$). As
explained in the Appendix \ref{appen:link}, the link between these fluid quantities is%
\begin{equation}
\frac{\sigma^{2}}{2}=\frac{2}{d}\tau_{\mathbf{U}}k_{T}=\frac{2}{d}\nu_{T}.\label{def sigma}%
\end{equation}

The first main result of our work is that we derive the following
Smoluchowski-type system for the particle densities of masses $m=1,2,...$
\begin{align}
&  \frac{\partial f_{m}\left(  t,\mathbf{x},\mathbf{v}\right)  }{\partial
t}+\mathbf{v\cdot\nabla}_{x}f_{m}\left(  t,\mathbf{x},\mathbf{v}\right)
-\gamma_{m}\operatorname{div}_{v}\left(  \mathbf{v}f_{m}\left(  t,\mathbf{x}%
,\mathbf{v}\right)  \right)  \label{SCeq}\\
&  -\frac{\gamma_{m}^{2}\sigma^{2}}{2}\Delta_{v}f_{m}\left(  t,\mathbf{x}%
,\mathbf{v}\right)  =\left(  \mathcal{Q}_{m}^{+}-\mathcal{Q}_{m}^{-}\right)
(\mathbf{f},\mathbf{f})(t,\mathbf{x},\mathbf{v})\nonumber
\end{align}
where $\mathbf{f}:=(f_{1},f_{2},...)$, $\mathbf{x}\in\mathbb{T}^{d}$ (the
$d$-dimensional torus),$\,\mathbf{v}\in\mathbb{R}^{d}$ and the collision
kernels are given by%
\begin{align}
\mathcal{Q}_{m}^{+}(\mathbf{f},\mathbf{f})(t,\mathbf{x},\mathbf{v}) &
:=\sum_{n=1}^{m-1}\iint_{\{n\mathbf{v}^{\prime}+(m-n)\mathbf{v}^{\prime\prime
}=m\mathbf{v}\}}s_{n,m-n}\label{eq:collision}\\
\cdot &  |\mathbf{v}^{\prime}-\mathbf{v}^{\prime\prime}|f_{n}(t,\mathbf{x}%
,\mathbf{v}^{\prime})f_{m-n}(t,\mathbf{x},\mathbf{v}^{\prime\prime
})d\mathbf{v}^{\prime}d\mathbf{v}^{\prime\prime},\nonumber\\
\mathcal{Q}_{m}^{-}(\mathbf{f},\mathbf{f})(t,\mathbf{x},\mathbf{v}) &
:=2f_{m}\left(  t,\mathbf{x},\mathbf{v}\right)  \sum_{n=1}^{\infty}\int
s_{n,m}\nonumber\\
&  \quad\cdot|\mathbf{v}-\mathbf{v}^{\prime}|f_{n}\left(  t,\mathbf{x}%
,\mathbf{v}^{\prime}\right)  d\mathbf{v}^{\prime}%
\end{align}
with $s_{n,m}$ defined in \eqref{surface-factor} below. 

This equation proposes a change of viewpoint. In previous works, the central
problem was determining the correct collision kernel which takes into account
the fact that the fluid is turbulent. Here we use the original collision
kernel depending on the relative velocity $\left\vert \mathbf{v}%
-\mathbf{v}^{\prime}\right\vert $, without modifying its coefficients, but
incorporate the presence of a small-scale turbulent background by adding the
dissipative operator in the velocity variable. Collision and aggregation is
not due to a stronger collision kernel, in this model, but to the
spread-in-$\mathbf{v}$ of densities, produced by the additional diffusion
term. 

We explain the derivation of this Smoluchowski-type system in Sections \ref{sec:stoc} and
\ref{sec:det} and in the Appendix \ref{appen-particle}. This derivation is heuristic but reasonable in analogy
with rigorous results proved recently for other models \cite{FlaGaleLuoJEE,
FlaGaleLuoPTRSA, Galeati}. From the viewpoint of the Physical validity of the
result, let us stress that the rigorous proof would require very small
$\tau_{\mathbf{U}}$, with $\gamma_{m}$ having a finite limit. Therefore $St$
must be large. 

We analyze this new model both using approximate analytical computations and
numerically. In Section \ref{sec:rel-vel} we prove, up to some approximation, the formula%
\begin{equation}
R_{m_{1},m_{2}}=\frac{2}{\sqrt{\pi}}\sqrt{\gamma_{m_{1}}+\gamma_{m_{2}}}\sigma=\frac{4}{\sqrt{3\pi}}\sqrt
{\frac{k_{T}}{St_{m_{1}}}+\frac{k_{T}}{St_{m_{2}}}},\label{average velocity}%
\end{equation}
in the physical dimension $d=3$. In the large $St$ regime, which is the regime of validity of our results, this
formula confirms known results (see the discussion in \cite{wilkinson2006caustic}) and it is known as the gas-kinetic model, after
\cite{abrahamson1975collision}. Let us notice that it is obtained without any
use of dimensional analysis; it is derived from basic equations, except for
the stochastic model of the turbulent fluid. It is not immediately clear,
however, if we may modify our approach to incorporate the concentration
effects related to singularities described in \cite{Falkovich, Mehlig, wilkinson2006caustic}.

In Section \ref{sec:numerical}, finally, we investigate numerically the Smoluchowski
equations, quantifying in various ways the efficiency of aggregation of the
turbulence model.

\section{The microscopic model}\label{sec:micro}

The model used below will be of Smoluchowski type with random transport.
However, the description of its microscopic origin may help. Call
$D\subset\mathbb{R}^{d}$, $d=1,2,3$, the space domain of the system, occupied
by the fluid and by small rain droplets. The number $\mathcal{N}(t)$ of
droplets changes in time due to coalescence. Droplet motion is described in a
Newtonian way by position and velocity $\left(  \mathbf{x}^{i}\left(
t\right)  ,\mathbf{v}^{i}\left(  t\right)  \right)  $, $i=1,...,\mathcal{N}%
(t)$. Droplets have masses $m^{i}\left(  t\right)  $ taking values in the
positive integers $\{1,2,...\}$. 
During the intertime between a collision and the next one, the motion is given by 
\begin{align*}
\frac{d\mathbf{x}^i}{dt}=\mathbf{v}^i,\qquad\frac{d\mathbf{v}^i}{dt}=\gamma_{m_i}\left(
\mathbf{U}\left(  t,\mathbf{x}^i\right)  -\mathbf{v}^i\right)
\end{align*}
where $\mathbf{U}\left(  t,\mathbf{x}\right)  $ is the fluid velocity; we
adopt a Stokes law for the particle-fluid interaction and denote by
\begin{align*}
\gamma_{m_i}=\alpha (m^i)^{\left(  1-d\right)  /d}
\end{align*}
the damping rate, $\alpha$ a
positive constant (including the dynamic viscosity coefficient of the fluid), and
the term $\left(  m^{i}\right)  ^{1/d}$ playing the role of the radius of the particle.

The rule of coalescence is crucial, see \cite{Devenish, Falkovich, Grabowski, Pumir, Saffman}. There are two typical
mathematical models: one is based on deterministic coalescence, the other on
probability rates. The first one is easier to describe: when two particles
meet, they become a new single particle with mass given by the sum of the
masses and momentum given by conservation of momentum. For mathematical
investigation of the macroscopic limit, this scheme is usually more difficult.
Easier is thinking in terms or \textit{rate of coalescence}: when two
particles are below a certain small distance one from the other, they have a
certain probability per unit of time to become a new single particle, with the
mass and momentum law as above. The kernels in Smoluchowski equations are the
macroscopic footprint of rates.

The model based on rates has a flaw precisely in connection with the
turbulence background we want to investigate here. Since coalescence happens
due to a probability per unit of time, if the time spent by two particles, at
the prescribed distance of potential coalescence, is small, the probability
that their encounter leads to coalescence is smaller. This is in sharp
contrast with the deterministic model where coalescence always happens, at a
certain distance, independently of the time spent nearby. In other words, in
the model based on rates, {without employing an approximating strategy to compute terminal velocity}, coalescence is facilitated by slow motion, which is
false in practice and goes in the opposite direction of understanding whether
turbulence enhances coalescence.

To avoid this bias towards slow motion, of say particles $i$ and $j$, and leave velocity as a studied attribute of the system, we
{maintain} in their coalescence rate the factor $|\mathbf{v}^{i}-\mathbf{v}^{j}|$. This factor multiplied by the time spent nearby is constant, on average, hence the probability of coalescence is roughly constant.

Finally, since the probability of coalescence should depend on the particle
surface, main factor involved in the collision, we multiple the rate  by the
surface factor
\begin{align}\label{surface-factor}
s_{m^{i},m^{j}}  =\left(  \left(  m^{i}\right)  ^{1/d}+\left(
m^{j}\right)  ^{1/d}\right)  ^{d-1}.
\end{align}
Hence, summarising, in our work the adopted point of view is consistent with the case of hydrodynamic 
motion, as in e.g. \cite{Falkovich, Mehlig}, where the coagulation kernel is 
\begin{align}\label{efficiency}
 E(i, j)s_{m^i, m^j}|\mathbf{v}^{i}-\mathbf{v}^{j}|,
\end{align}
and the scalar $E(i,j)$ can be regarded as collision efficiency
between real droplets $i$ and $j$. For simplicity, we set $E(i,j)=1$ in our phenomenological study.

\section{The Smoluchowski-type model}\label{sec:smo-type}

A rigorous study of the link between the microscopic model and the macroscopic
one is under investigation, following \cite{FlandHuang1, FlandHuang2, Hammond, Papini} where similar models have been already
treated. However, following the mean field paradigm we may safely choose the
following macroscopic model as a good one for the density evolution.

Denote by $f_{m}\left(  t,\mathbf{x},\mathbf{v}\right)  $, $m=1,2,...$,
the density of droplets of mass $m$ at position $\mathbf{x}\in D$ having
velocity $\mathbf{v\in}\mathbb{R}^{d}$ . Then (dropping the time variable) the
density satisfies
\begin{align}\label{spde-krai}
&\frac{\partial f_{m}\left(  \mathbf{x},\mathbf{v}\right)  }{\partial
t}+\operatorname{div}_{x}\left(  \mathbf{v}f_{m}\left(  \mathbf{x}%
,\mathbf{v}\right)  \right) \nonumber \\
&+\gamma_m\operatorname{div}_{v}\left(  \left(
\mathbf{U}\left(  t,\mathbf{x}\right)  -\mathbf{v}\right)  f_{m}\left(
\mathbf{x},\mathbf{v}\right)  \right)  =\mathcal{Q}_{m}^{+}-\mathcal{Q}%
_{m}^{-}%
\end{align}
where $\gamma_m = \alpha m^{(1-d)/d}$, and $\mathcal{Q}_{m}^{+}$ and
$\mathcal{Q}_{m}^{-}$ are the two collision terms as given in \eqref{eq:collision}.
Crucial is the kernel $|\mathbf{v}^{\prime}-\mathbf{v}^{\prime\prime}|$, as
described above. The first collision term describes the amount of new
particles of mass $m$ created by collision of smaller ones, with the momentum
conservation rule 
\begin{align}\label{conserv-momentum}
n\mathbf{v}^{\prime}+(m-n)\mathbf{v}^{\prime\prime
}=m\mathbf{v}. 
\end{align}
The second collision term gives us the percentage of the
density $f_{m}\left(  \mathbf{x},\mathbf{v}\right)  $ of particles of mass $m$
which disappears by coalescence into larger particles.

In the next section, we explain how this model can be studied using techniques from passive scalars, thus obtaining in \eqref{SCeq} a 
simplified 
coagulation equation in which the velocity of the particles 
is still a driving component of the coalescence process. We postpone to the Appendix \ref{appen-particle} (see also \cite{FHP})
for a more rigourous heuristic of the scaling limit from a coagulating microscopic particle system subjected to a common noise, to 
a stochastic partial differential equation (\abbr{SPDE}), that eventually gives
rise to the \abbr{PDE} \eqref{SCeq}. Although it is not yet fully rigorous, we believe that it justifies the interest of
this equation. The eddy diffusion now occurs in the velocity variable.

\section{Stochastic model of turbulent velocity field}\label{sec:stoc}

Similarly to a large body of simplified modeling of passive scalars, we
consider a model of velocity fluid which is delta-correlated in time, namely a
white noise with suitable space dependence. We may write%
\begin{align}\label{white-noise}
\mathbf{U}\left(  t,\mathbf{x}\right)  dt=\sum_{k\in K}\mathbf{\sigma}%
_{k}\left(  \mathbf{x}\right)  dW_{t}^{k}%
\end{align}
where $\mathbf{\sigma}_{k}\left(  \mathbf{x}\right)  $ are smooth divergence
free deterministic vector fields on $D$ and $W_{t}^{k}$ are independent
one-dimensional Brownian motions; $K$ is a finite index set (or countable,
with some care on summability assumptions). In this case the term
$\gamma_m\mathbf{U}\left(  t,\mathbf{x}\right)  \cdot\nabla_{v}f_{m}\left(
\mathbf{x},\mathbf{v}\right)  $ must be interpreted as a Stratonovich integral
(still written here in differential form for sake of clarity)%
\[
\gamma_m\sum_{k\in K}\mathbf{\sigma}_{k}\left(  \mathbf{x}\right)  \cdot\nabla
_{v}f_{m}\left(  \mathbf{x},\mathbf{v}\right)  \circ dW_{t}^{k}.
\]
By the rules of stochastic calculus, it is given by an It\^{o}-Stratonovich
corrector plus an It\^{o} integral; precisely, the previous term is given by
\[
-\frac{\gamma^2_m}{2}\sum_{k\in K}\mathbf{\sigma}_{k}\left(  \mathbf{x}\right)
\cdot\nabla_{v}\left(  \mathbf{\sigma}_{k}\left(  \mathbf{x}\right)
\cdot\nabla_{v}f_{m}\left(  \mathbf{x},\mathbf{v}\right)  \right)
dt+dL\left(  t,\mathbf{x},\mathbf{v}\right)
\]
where $L\left(  t,\mathbf{x}, \bv\right)  $ is a (local) martingale, the It\^{o}
term. The It\^{o}-Stratonovich corrector takes also the form%
\[
-\frac{\gamma^2_m}{2}\operatorname{div}_{v}\left(  C \left(  \mathbf{x}%
,\mathbf{x}\right)  \nabla_{v}f_{m}\left(  \mathbf{x},\mathbf{v}\right)
\right)  dt
\]
where $C \left(  \mathbf{x},\mathbf{y}\right)  $ is the matrix-valued function
given by the space-covariance function of the noise%
\begin{align}\label{noise-cov}
C \left(  \mathbf{x},\mathbf{y}\right)  =\sum_{k\in K}\mathbf{\sigma}%
_{k}\left(  \mathbf{x}\right)  \otimes\mathbf{\sigma}_{k}\left(
\mathbf{y}\right)  .
\end{align}
Summarizing, the stochastic model, in It\^{o} form, is%
\begin{align}\label{stoc-model}
&  df_{m}\left(  \mathbf{x},\mathbf{v}\right)  +\left(  \mathbf{v\cdot\nabla
}_{x}f_{m}\left(  \mathbf{x},\mathbf{v}\right)  -\gamma_m\operatorname{div}%
_{v}\left(  \mathbf{v}f_{m}\left(  \mathbf{x},\mathbf{v}\right)  \right)
\right)  dt\nonumber\\
&  -\frac{\gamma^2_m}{2}\operatorname{div}_{v}\left(  C \left(  \mathbf{x}%
,\mathbf{x}\right)  \nabla_{v}f_{m}\left(  \mathbf{x},\mathbf{v}\right)
\right)  dt\nonumber\\
&  =\left(  \mathcal{Q}_{m}^{+}-\mathcal{Q}_{m}^{-}\right)  dt-dL\left(
t,\mathbf{x},\mathbf{v}\right)  .
\end{align}
Also for later reference, let us mention an example of noise, introduced by R.
Kraichnan \cite{Kr, Krai}, relevant to our analysis. For the sake of
simplicity of exposition, assume we are in full space $\mathbb{R}^{d}$, but
modifications in other geometries are possible. Its covariance function is
space-homogeneous, $C \left(  \bx,\by\right)  = C \left(  \bx-\by\right)  $, with the
form%
\begin{align}\label{krai-cov}
C \left( \bz\right)  =\sigma^{2}k_{0}^{\zeta}\int_{k_{0}\leq\left\vert
\bk\right\vert <k_{1}}\frac{1}{\left\vert \bk\right\vert ^{d+\zeta}}e^{i\bk\cdot
\bz}\left(  I-\frac{\bk\otimes \bk}{\left\vert \bk\right\vert ^{2}}\right)
d\bk\mathbf{.}%
\end{align}
The case $\zeta>0$ includes Kolmogorov 41 case $\zeta=4/3$. In this case, take
$k_{1}=+\infty$. Then 
\[
C \left(  \mathbf 0\right)  =A\sigma^{2} \]
where the constant $A$
is given by 
\[
\int_{1\leq\left\vert \bk\right\vert <\infty}\frac{1}{\left\vert
\bk\right\vert ^{d+\zeta}}\left(  I-\frac{\bk\otimes \bk}{\left\vert \bk\right\vert
^{2}}\right)  d\bk.
\] 

\section{The deterministic scaling limit}\label{sec:det}

Following \cite{FlaGaleLuoJEE, FlaGaleLuoPTRSA, Galeati}, we may
consider small-scale turbulent velocity fields depending on a scaling
parameter and take their scaling limit. In the case of Kraichnan model above,
choose
\[
k_{0}=k_{0}^{N}\rightarrow\infty
\]
The result $C \left( \mathbf 0\right)  =A\sigma^{2}I_d$ is independent of $N$, so that
the It\^{o}-Stratonovich corrector becomes equal to (without loss of generality we set $A=1$)
\[
\frac{1}{2}\gamma_m^2\sigma^2\Delta_{v}%
f_{m}\left(  \mathbf{x},\mathbf{v}\right); 
\]
and simultaneously we may have
that the It\^{o} term goes to zero. The final equation is deterministic, and
precisely given by%
\begin{align*}
&  \frac{\partial f_{m}\left(  \mathbf{x},\mathbf{v}\right)  }{\partial
t}+\mathbf{v\cdot\nabla}_{x}f_{m}\left(  \mathbf{x},\mathbf{v}\right)
-\gamma_m\operatorname{div}_{v}\left(  \mathbf{v}f_{m}\left(  \mathbf{x}%
,\mathbf{v}\right)  \right) \\
&  -\frac{\gamma^2_m\sigma^{2}}{2}\Delta_{v}f_{m}\left(  \mathbf{x}%
,\mathbf{v}\right)  =\mathcal{Q}_{m}^{+}-\mathcal{Q}_{m}^{-}.
\end{align*}

Now, for sake of numerical simplicity, we assume that all densities are
uniform in $\mathbf{x}$. Then we have%

\begin{equation}
\boxed{
  \!\begin{aligned}
\frac{\partial f_{m}\left(  \mathbf{v}\right)  }{\partial t}%
-&\gamma_m\operatorname{div}_{v}\left(  \mathbf{v}f_{m}\left(  \mathbf{v}\right)
\right)  -\frac{\gamma^2_m\sigma^{2}}{2}\Delta_{v}f_{m}\left(  \mathbf{v}%
\right) \\
&=(\mathcal{Q}_{m}^{+}-\mathcal{Q}_{m}^{-})(\mathbf f, \mathbf f)(\bv) \label{finaleq}  \end{aligned}
}
\end{equation}
where now the collision term $\mathcal{Q}_{m}^{+}-\mathcal{Q}_{m}^{-}$
includes only functions of $\bv$. This is our final equation for the density of
droplets. It is parametrized by $\sigma^2$, the intensity of noise covariance which, in the approximation of this white noise model, corresponds to the concept of {\it turbulence kinetic energy}, cf. \cite{dupuy2019effect}. Even though \eqref{finaleq} is of variable $\bv$ only, it is fundamentally different from a Smoluchowski equation with only $\bx$ variable, due to the presence of velocity difference $|\bv-\bv'|$ in the nonlinearity. This term is the source that turns diffusion enhancement into coagulation enhancement.

\section{Formula for the average relative velocity}\label{sec:rel-vel}

In order to approximate analytically the average value $\left\langle
\left\vert \mathbf{v}_1-\mathbf{v}_2\right\vert \right\rangle $ we adopt
the mean field viewpoint of Smoluchowski equations, where particles are
independent. Therefore, if $p_{m}\left(  \mathbf{v}\right)  $ is the
probability density of velocity of mass $m$, we have%
\begin{equation}
R_{m_1,m_2}=\iint\left\vert \mathbf{v}_1-\mathbf{v}_2\right\vert
p_{m_1}\left(  \mathbf{v}_1\right)  p_{m_2}\left(  \mathbf{v}_2\right)  d\mathbf{v}_1d\mathbf{v}_2.\label{limitDens}
\end{equation}
The natural choice of $p_{m}\left(  \mathbf{v}\right)  $ is the normalized
density $f_{m}\left(  \mathbf{v}\right)  /\int f_{m}\left(  \mathbf{w}\right)
d\mathbf{w}$ where $f_{m}\left(  \mathbf{v}\right)  $ is a solution of
Smoluchowski equation. However, we have to avoid a dependence on the initial
conditions. We make the following heuristic argument. In the Smoluchowski
system, the linear terms
\[
\gamma_m\operatorname{div}_{v}\left(  \mathbf{v}f_{m}\left(  \mathbf{v}\right)
\right)  +\frac{\gamma^2_m\sigma^{2}}{2}\Delta_{v}f_{m}\left(  \mathbf{v}%
\right)
\]
are associated with the transient phase which moves the initial distribution
towards a certain limit shape. Simultaneously and afterwards, the nonlinear
terms shift mass from lower to higher levels, but their impact on the
modification of shape is minor. Therefore we take, as $p_{m}\left(
\mathbf{v}\right)  $ the invariant distribution of the linear part, which is a
centered Gaussian with covariance matrix $\frac{1}{2}\gamma_m\sigma^{2}I_d$
($I_d$ is the identity matrix):%
\[
p_{m}\sim N\left(  0,\frac{1}{2}\gamma_m\sigma^{2}I_d\right)  .
\]
The difference of two independent centered Gaussians, with covariances
$\frac{1}{2}\gamma_{m_1}\sigma^{2}I_d$ and $\frac{1}{2}\gamma_{m_2}\sigma^{2}I_d$ is a centered Gaussian
with covariance $\frac{1}{2}\left(  \gamma_{m_1}+\gamma_{m_2}\right)  \sigma^{2}I_d$.
Therefore the random quantity $\mathbf{v}_1-\mathbf{v}_2$ has this law.
By properties of Gaussians, 
\[
\bv_1-\bv_2 \overset{(d)}{=} \sqrt{\frac{1}{2}\left(  \gamma_{m_1}+\gamma_{m_2}\right)  }\sigma \bZ 
\]
where $\bZ$ is distributed as $N(0, I_d)$, and 
\[
\langle |\bZ|\rangle =\sqrt{2}\frac{\Gamma(\frac{d+1}{2})}{\Gamma(\frac{d}{2})}
\]
since $|\bZ|$ has a Chi distribition with parameter $d$. 
Thus we have%
\[
R_{m_1,m_2}=\langle |\bv_1-\bv_2|\rangle=\frac{\Gamma(\frac{d+1}{2})}{\Gamma(\frac{d}{2})}\sqrt{\gamma_{m_1}+\gamma_{m_2}}\sigma.
\]
By \eqref{def sigma}, $\sigma^2=\frac{4}{d}\tau_{\bU}k_T$ and taking $d=3$, $\Gamma(2)=1$, $\Gamma(\frac{3}{2})=\frac{\sqrt{\pi}}{2}$, we arrive at   
\begin{align*}
R_{m_1,m_2}&=\sqrt{\frac{4}{3}}\frac{2}{\sqrt{\pi}}\sqrt{(\gamma_{m_1}\tau_{\bU}+\gamma_{m_2}\tau_{\bU})k_T}\\
&=\frac{4}{\sqrt{3\pi}}\sqrt
{\frac{k_{T}}{St_{m_{1}}}+\frac{k_{T}}{St_{m_{2}}}},
\end{align*}
as announced in \eqref{average velocity}.

Up to the multiplicative constant, this also agrees with the formula obtained by Abrahamson \cite{abrahamson1975collision}. Indeed, in \cite{abrahamson1975collision} the energy dissipation rate $\ep\sim k_T/\tau_\bU$, clear from the energy balance of Navier-Stokes equation since all three quantities correspond
to the turbulent fluid:
\[
\frac{\partial}{\partial t}\left(\frac{1}{2}|\bU|^2\right)=-\ep +\text{other terms}.
\]

\section{Numerical results}\label{sec:numerical}
For the convenience of numerical simulations, we consider from now on only finitely many mass levels. That is, we truncate \eqref{finaleq} into a finite system of \abbr{PDE}-s whose solution is $(f_1,f_2,...,f_M)$, for some integer $M$. This amounts to replacing the $\sum_{n=1}^\infty$ in the loss term $\cQ_m^-$ \eqref{eq:collision} by $\sum_{n=1}^M$, with everything else unchanged. Correspondingly, in the particle system \eqref{particle-fluid}, each particle's mass is restricted to $m_i\in\{1,2,...,M\}$. The interpretation is that when the mass of a rain droplet exceeds the threshold $M$, it falls down and hence exits the system. 

To understand the effect of the turbulent velocity field on coagulation, we identify  and build on a key quantity, $\cM_1^\sigma(t)$ below, which is essentially the  first moment of the mass in the system at time $t$. Since $M<\infty$ in the truncated model, eventually all masses leave the system, hence we measure the efficiency of coagulation by looking at how fast this first moment decays in time, with respect to different values of $\sigma$.
In the last part of this section, using results on the total mass, we will build a procedure to estimate the mean Collision Rate (see section \ref{sec:rel-vel}), validating our theoretical results in simple settings.

\subsubsection{Total mass}
To this end, we define 
\begin{align}\label{totmass}
    \mathcal{M}^\sigma_1(t):=\sum_{m=1}^M m\int f_m(t,\bv)\, d\bv\, ,
\end{align}
which we also call ``total mass'' for simplicity. 
Analyzing the nonlinearity of our PDE, we notice that
\begin{align}\label{nonlin-dec}
\sum_{m=1}^M\int m(\mathcal Q^+_m-\mathcal Q^-_m) \, d\bv\leq 0, \quad \forall t
\end{align}
implying that $d\cM_1^\sigma(t)/dt\le 0$, that is, the function \eqref{totmass} is non-increasing in time. Moreover,
for the infinite system $M=\infty$,  
equality is achieved in \eqref{nonlin-dec}, hence we see  that the mass deficiency in the finite system is not lost at all and it is 
simply sent to higher order  ($>M$) of mass-type densities. 

Indeed, in view of the form of the negative part of  coagulation operator $\mathcal Q^-_m$, every coagulation at the level of $f_m, f_n$, with $m+n>M$, represents a decrease in mass that, ideally, increases the density $f_{m+n}$ that is outside of our system.
In particular, fixing $M<\infty$, in the framework of rain formations, is equivalent to saying that such a threshold represents the largest droplets that are falling outside of the cloud and do not interact any more with the system. As such $M=\infty$ is just the precise abstract setting in which no rainfall is present and serves as a limiting behavior for the single masse $m\in\mathbb{N}$, and as a right derivation of the conserved mass in the system as all: both for the falling particles and the ones remaining in the cloud.
Hence, the more and faster the quantity $\mathcal{M}_1^\sigma(t)$ decreases over time, the faster and richer the coagulation to higher mass-type is achieved.%

\subsubsection{Faster barrier exit time}
The second quantity we consider is closely linked to the enhanced coagulation due to 
turbulence that we will establish with the ``total mass" and gives more quantitative information. We will consider the same numerical setting as we will do above, and estimate a decay law that links the first time that the total mass $\cM_1^\sigma(t)$ drops below a certain level to the turbulence parameter $\sigma$. Specifically, let
\begin{align*}
m_0^T:=\inf_{t\in[0,T]}\mathcal{M}_1^0(t)
\end{align*}
and define a sequence of ``barrier exit times'' $(\tau_\sigma)_{\sigma\ge0}$
\begin{align}\label{barrier}
    \tau_\sigma&:=\inf\left\{t\ge0, \,\mathcal{M}_1^\sigma(t)\leq m_0^T\right\}\wedge T.
\end{align}
Since $t\mapsto\mathcal{M}_1^0(t)$ is decreasing, we have that $\tau_0=T$. 
Since $\cM_1^\sigma(t)$ is expected to decay faster as $\sigma$ increases, $\sigma\mapsto\tau_\sigma$ should be decreasing.
{\begin{figure}[h!]
\includegraphics[
height=0.45\textwidth,
width=0.45\textwidth
]%
{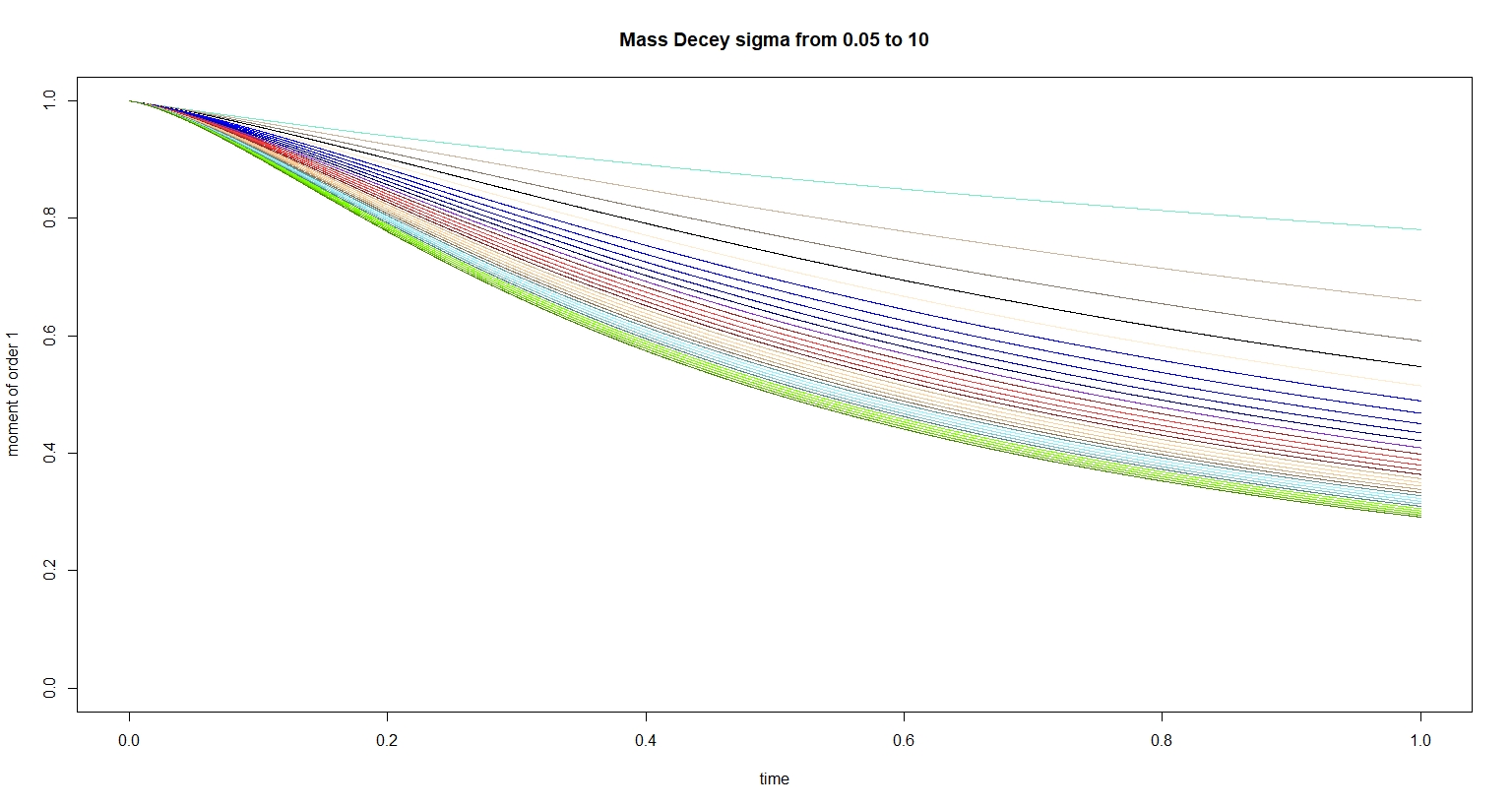}
\caption{$M=1$; Decay of $\mathcal{M}_1^\sigma(t)$ for $t\in[0,1]$, with maximal mass level $M=1$, initial density $f_1(0, \bv)$ of mass $m=1$  concentrated on the set $\bv\in[-1/2,1/2]$. The parameter $\sigma^2$ ranges from a sample in the set $0.05$ to $10$ (around $30$ points). A visible increase in coagulation is present at the increase of $\sigma^2$.}
\label{fig1}
\end{figure}
}

\subsection{On a limiting behavior: $M=1$}\label{sec:M1}
 {\begin{figure*}[!htbp]
\includegraphics[
height=0.35\textwidth,
width=0.6\textwidth
]%
{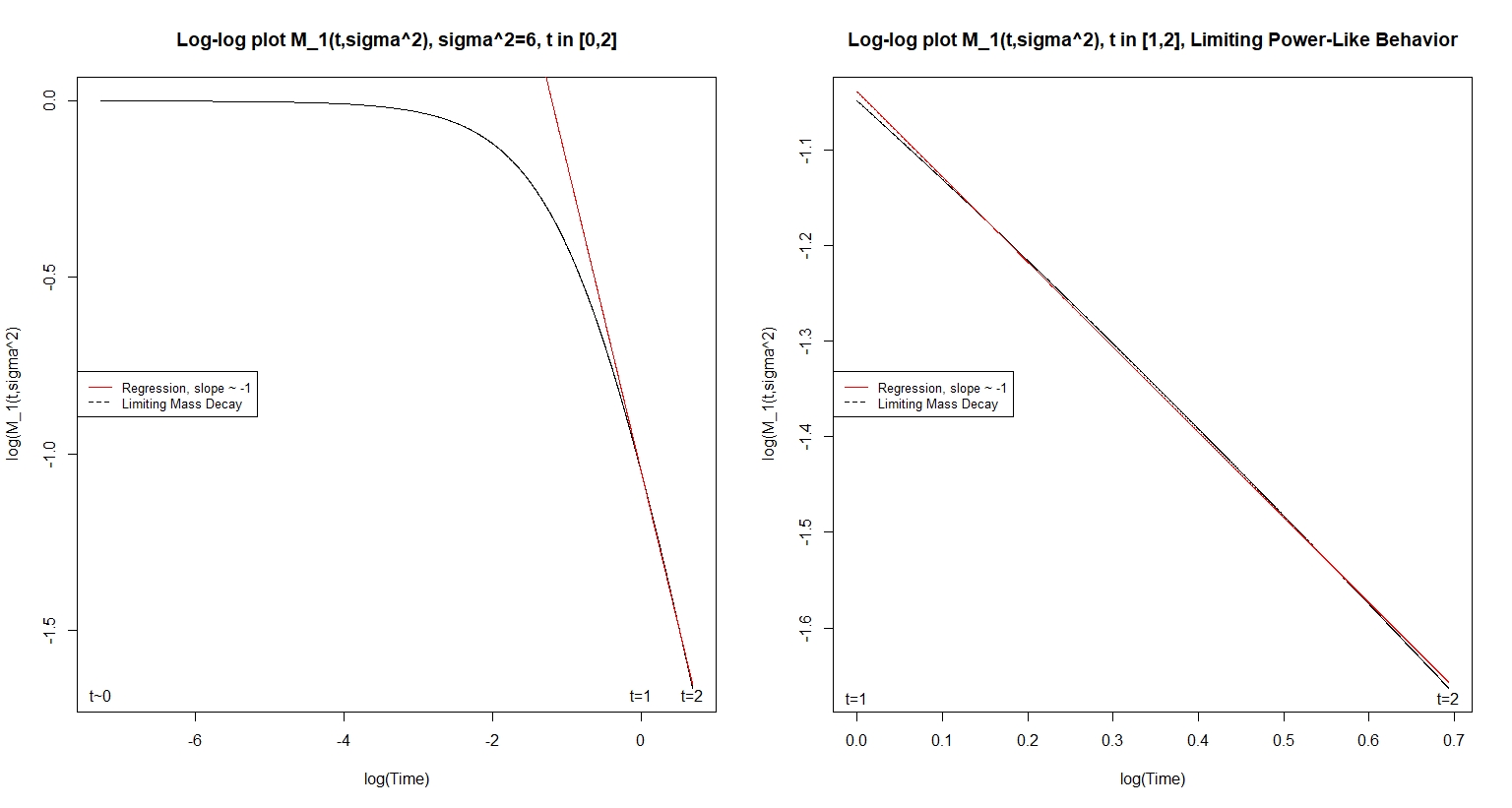}
\caption{$M=1$; On the left, a plot of $\log\left(\cM_1^\sigma(t)\right)$ versus $\log (t)$ in the time window $[0,2]$ at fixed $\sigma^2=6$, and on the right a close-up in the time window $[1,2]$, suggest that $t\mapsto \cM_1^\sigma(t)$ is of inverse power $1$. However, for small time, the dependence is different and could represent a transient behavior.}
\label{fig:inverse-power-1}
\end{figure*}}

{\begin{figure*}[!htbp]
   
        \includegraphics[
height=0.35\textwidth,
width=0.6\textwidth
]%
{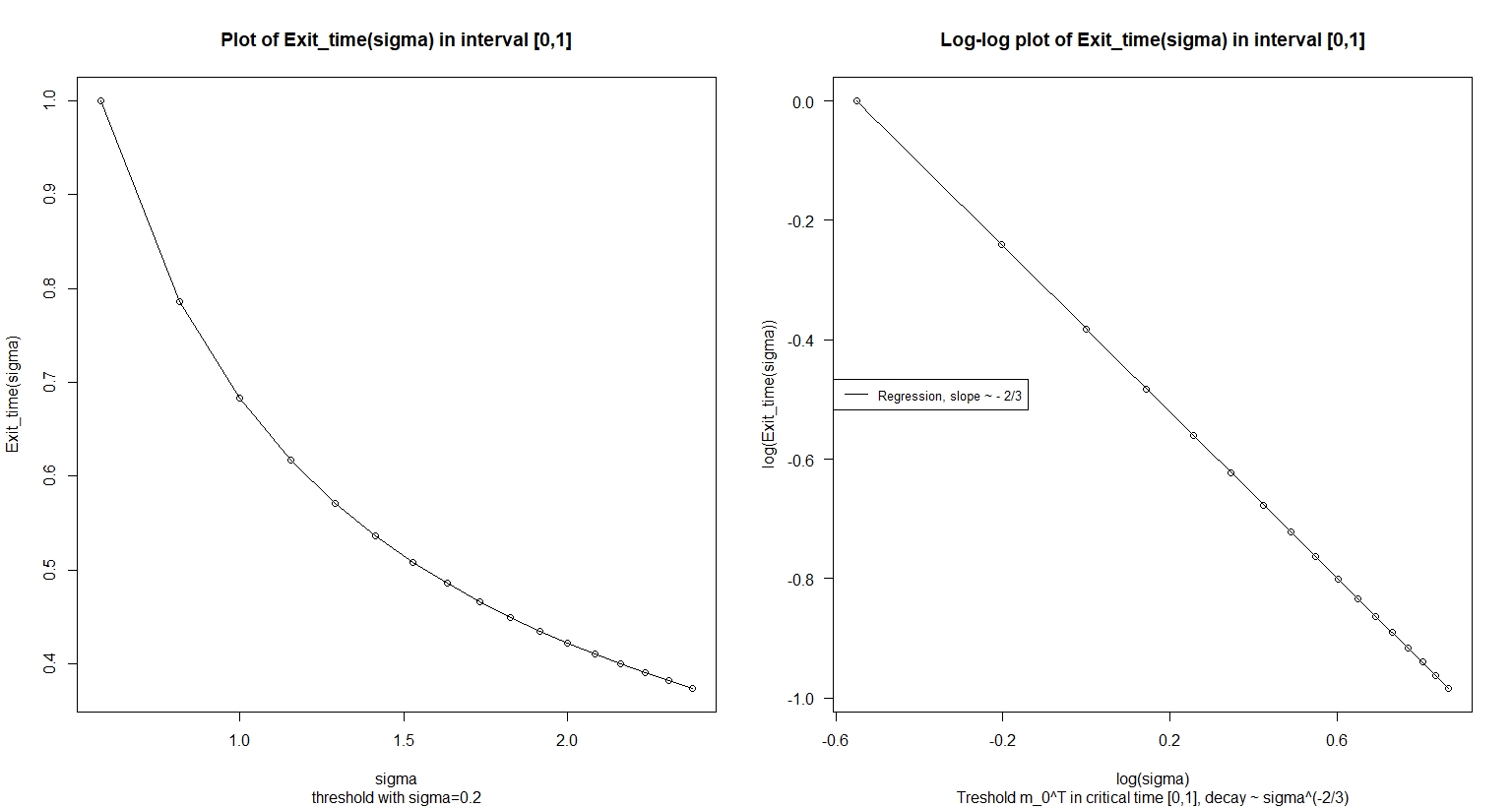}
         \caption{$M=1$; A plot of the barrier exit time $\tau_\sigma$ with respect to the turbulence parameter $\sigma$, and the corresponding log-log regression in the time window $[0,1]$ yields $\tau_\sigma \propto \sigma^{-2/3}$. }
         \label{fig:loglog}

     \end{figure*}}

{\begin{figure*}[!htbp]
   
        \includegraphics[
height=0.35\textwidth,
width=0.6\textwidth
]%
{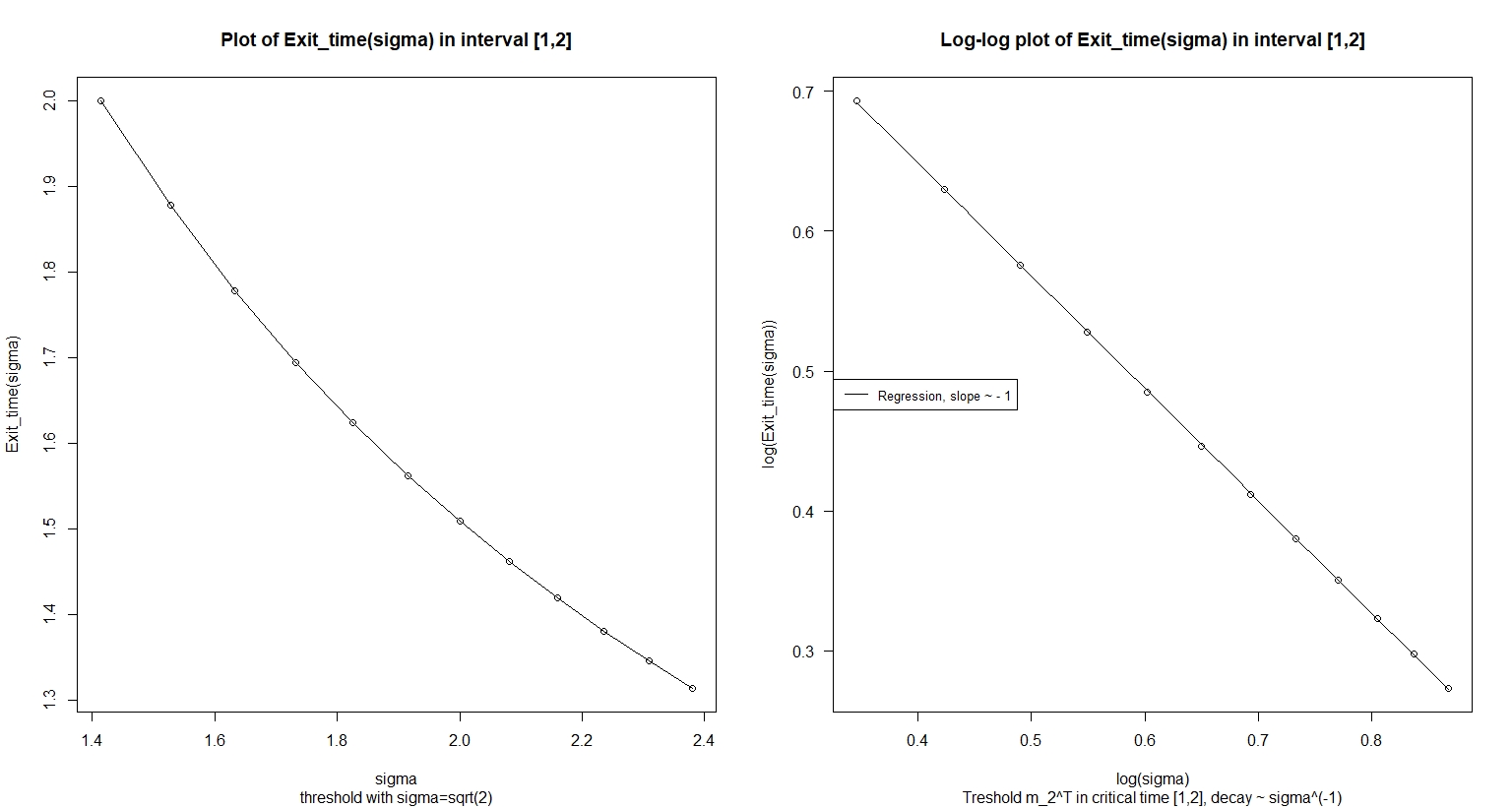}
         \caption{$M=1$; A plot of the barrier exit time $\tau_\sigma$ with respect to the turbulence parameter $\sigma$, and the corresponding log-log regression in the time window $[0,2]$, taking into consideration only those exit times in the interval $[1,2]$, yields $\tau_\sigma \propto\sigma^{-1}$.  }
         \label{fig:loglog-2}

\end{figure*}}

We perform a numerical simulation of the system \eqref{finaleq} for 
dimension $d=1$, maximal mass level $M=1$ and time window $[0,2]$, with a semi-implicit method to compute its solutions. 
Thanks to the fast decay to zero as $|\bv|\to\infty$
of the solution  \cite{FHP}, we truncate the velocity variable in the range $\bv\in[-20,20]$ both for the numerical integration of the nonlinearity and for the total mass \eqref{totmass}. 

In Figure \ref{fig1}, we plot the function \eqref{totmass} for different values of the turbulence parameter $\sigma^2$ that range from $0.05$, that we refer to as the non-turbulent case, to $10$, which represents an intense eddy diffusivity. 
It shows a faster decay correlated to the increase of turbulence, 
and a speedup coagulation process.

For fixed $\sigma^2=6$ we performed a log-log plot in time window $[0, 2]$ as shown in Figure \ref{fig:inverse-power-1} that shows $t\mapsto \cM_1^\sigma(t)$ is of inverse power $1$, after a transient time period. 

We see from Figures \ref{fig:loglog} and \ref{fig:loglog-2} that the expected behavior on the barrier time is obtained, 
and the curve exhibits a power like decay, with an asymptotic limit to zero. In Figure \ref{fig:loglog}, we performed a log-log plot and regression taking $T=1$ and it yields $\tau_\sigma\propto\sigma^{-2/3}$ (here and in the sequel $\propto$ denotes proportional to), whereas the same analysis in Figure \ref{fig:loglog-2} taking $T=2$ and considering only those exit times that are in the interval $[1,2]$ yields $\tau_\sigma \propto\sigma^{-1}$. 

We conjecture that the function \eqref{totmass} can be expressed as (for $t$ suitably large, say $t>1$ in our simulations)
\begin{align}\label{self-similar-1}
\cM_1^\sigma(t)\sim \frac{1}{A_d(\sigma) t +{\cM^\sigma_1(0)}^{-1}},
\end{align}
for some function $A_d$ that depends on dimension $d$, and that $A_1(\sigma)\propto\sigma$. Here and in the sequel, $\sim$ denotes asymptotically for large $t$. 

A rough explanation of the numerical findings may be the following one, that
will be explored more closely in a future work, since - as shown below - our
understanding is still incomplete. When $M=1$, the density $f\left(
t,\mathbf{v}\right)  $ of the unique level $m=1$ satisfies the identity%

\[
\frac{d}{dt}\int f\left(  t,\mathbf{v}\right)  d\mathbf{v}=-\iint\left\vert
\mathbf{v}-\mathbf{v}^{\prime}\right\vert f\left(  t,\mathbf{v}\right)
f\left(  t,\mathbf{v}^{\prime}\right)  d\mathbf{v}d\mathbf{v}^{\prime}%
\]
because the differential terms cancel by integration by parts. Assume that, at
least after a transient time (confirmed by Figure \ref{fig:inverse-power-1}), up to a small
approximation,
\[
f\left(  t,\mathbf{v}\right)  \sim\alpha\left(  t\right)  f_{0}\left(
\mathbf{v}\right)
\]
namely the decay of $f\left(  t,\mathbf{v}\right)  $ is self-similar
\cite{Eggers}. Then (up to approximation) $\alpha^{\prime}=-\sigma_{0}%
\alpha^{2}$ where 
\[
\sigma_{0}=\iint\left\vert \mathbf{w}-\mathbf{w}%
^{\prime}\right\vert f_{0}\left(  \mathbf{w}\right)  f_{0}\left(
\mathbf{w}^{\prime}\right)  d\mathbf{w}d\mathbf{w}^{\prime} 
\]
is an average
variation of velocity under $f_{0}$, namely
\[
\alpha\left(  t\right)  \sim\frac{1}{\sigma_{0}t+C}%
\]
after an initial transient period. Moreover, speculating that the standard deviation
of $f_{0}$ should be of order $\sigma$ (since the dispersion produced by the
linear differential operator is proportional to $\sigma$), we expect that
$\sigma_{0}$ increases linearly with $\sigma$. The numerical results of
Figures \ref{fig:loglog} and \ref{fig:loglog-2} show that this looks the trend for sufficiently large time but
for a short time another power, $\sigma^{2/3}$, emerges, that should be
understood. As for the behavior in time, since this computation can be carried out for every $d>1$, when $M=1$, we believe that the decay in time is dimension-independent.


{\begin{figure}[h!]
\includegraphics[
height=0.5\textwidth,
width=0.5\textwidth
]%
{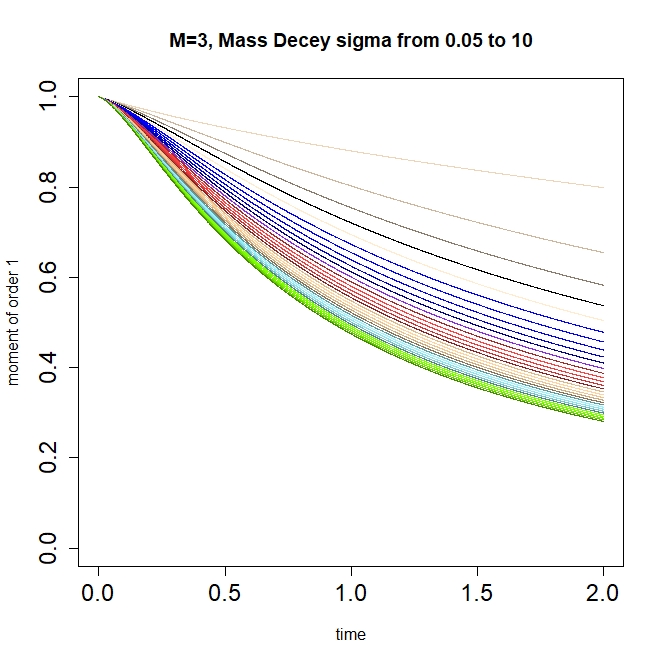}
\caption{$M=3$; Decay of $\mathcal{M}_1^\sigma(t)$ for $t\in[0,2]$, with maximal mass level $M=3$, initial density $f_1(0, \bv)$ of mass $m=1$  concentrated on the set $\bv\in[-1/2,1/2]$, $f_j(0, \bv)=0, j\neq1$. The parameter $\sigma^2$ ranges from a sample in the set $0.05$ to $10$ (around $30$ points). A visible increase in coagulation is present at the increase of $\sigma^2$.}
\label{fig1M3}
\end{figure}
}
{\begin{figure*}[!htbp]
   
        \includegraphics[
height=0.35\textwidth,
width=0.6\textwidth
]%
{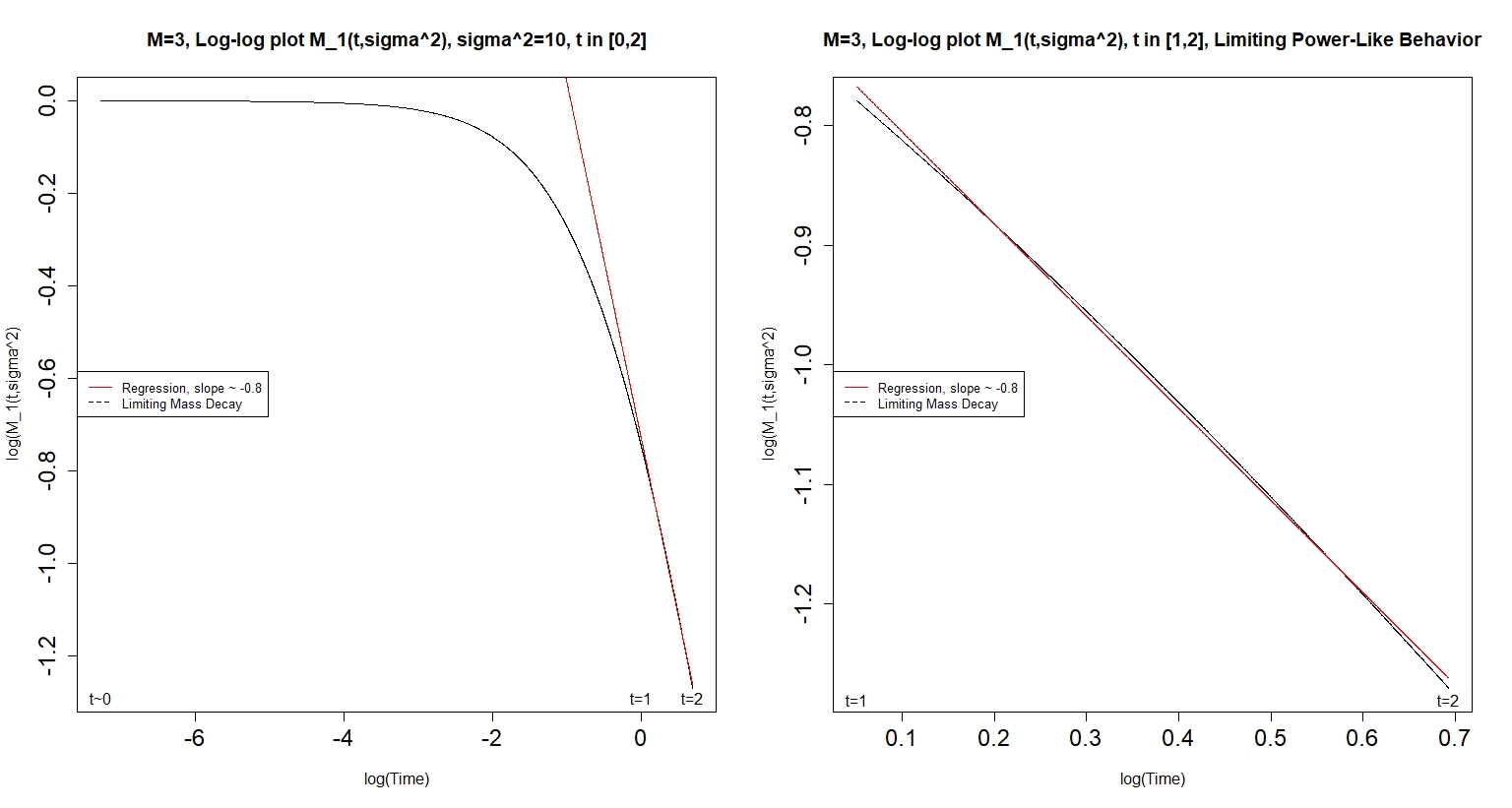}
         \caption{$M=3$; On the left, a plot of $\log\left(\cM_1^\sigma(t)\right)$ versus $\log (t)$ in the time window $[0,2]$ at fixed $\sigma^2=10$, and on the right a close-up in the time window $[1,2]$, suggests that $t\mapsto \cM_1^\sigma(t)$ is of inverse power $0.8$. However, for small time, the dependence is different and could represent a transient behavior.}
         \label{fig:loglogM3}
     
     \end{figure*}}
\subsection{Localized mass concentration: $M>1$}
When considering $M>1$, we can expect two natural settings to investigate: the one where initially all the mass is concentrated on the first level, i.e. $m=1$, and the one that follows the theoretical assumptions of \cite{Hammond,FHP,FlandHuang2}. Concerning the first setting, we perform a numerical simulation of the system \eqref{finaleq} for 
dimension $d=1$, maximal mass level $M=3$ and time window $[0,2]$.

In Figure \ref{fig1M3}, we plot the function \eqref{totmass} for different values of the turbulence parameter $\sigma^2$ that ranges from $0.05$, that we refer to as the non-turbulent case, to $10$, which represents an intense eddy diffusivity. 
As in the case of $M=1$, it shows a faster decay correlated to the increase of turbulence, 
and a speedup coagulation process.

For fixed $\sigma^2=10$, we perform a log-log plot in time window $[0, 2]$ as shown in Figure \ref{fig:loglogM3} that shows $t\mapsto \cM_1^\sigma(t)$ is of inverse power approximately of $0.8$, after a transient time period. Thus, we see a difference in the behavior of the ``total mass" when $M$ increases: this is not unexpected when all the initial mass is concentrated in the first layer $m=1$. In fact, analyzing the coagulation operator (\ref{eq:collision}), we see that $\mathcal{Q}^+_m$ is responsable for the generation of bigger particles in higher mass-levels and it is dominant when all the mass of the system is selected as a single type. Therefore, for a transient period, we see an increase in mass for $m\neq1$ and as such a slower decay of $\mathcal{M}_1^\sigma(t)$, the total mass.

For this reason, as shown in Figure \ref{fig:inverse-power-1-M3}, we study the decay of the single mass $m\in\{1,2,3\}$, where analogous to \eqref{totmass}, the single mass at level $m=k$ is defined as  
\begin{align}\label{single-mass}
\cM_1^\sigma(t) |_{m=k}:=k\int f_k(t,\bv)\, d\bv.
\end{align}
The figure shows the regression curves plotted with dashed lines. As in Figure \ref{fig:inverse-power-1}, for $m=1$ we maintain a relation of inverse power in time, approximately of $1$, after a transient time period.  As a further exploration, we see from Figure \ref{fig:loglogM3m1} that the same behavior is present, 
and the curve exhibits a power like decay, with an asymptotic limit to zero.
    
 {\begin{figure}[!htbp]
\includegraphics[
height=0.45\textwidth,
width=0.45\textwidth
]%
{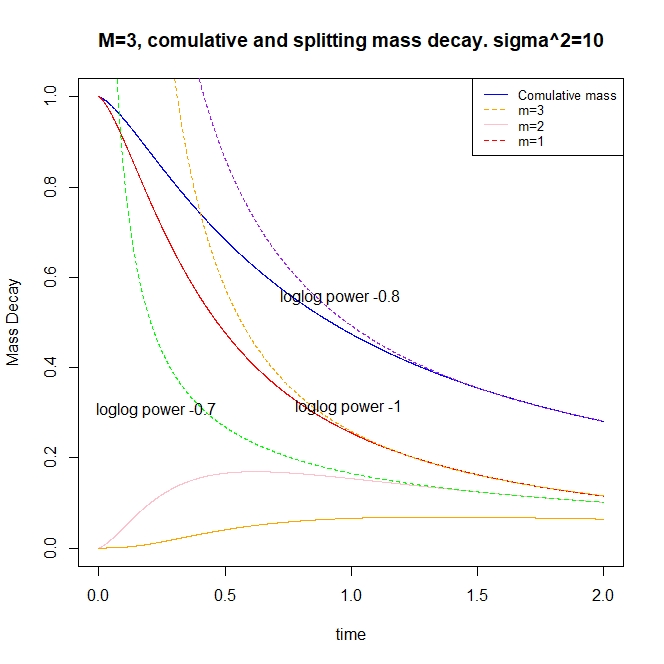}
\caption{$M=3$; Decay of $\mathcal{M}_1^\sigma(t)$ for the total mass, and the single behavior $\cM_1^\sigma(t)|_{m=k}$ of each lever $k\in\{1,2,3\}$ in the case $\sigma^2=10$. With the dashed lines, one can see the expected limiting behaviors of each curve and their relative power. This suggest a log-logistic behavior of the full system with $M<\infty$.}
\label{fig:inverse-power-1-M3}
\end{figure}}

Concerning the behavior of the barrier time, we see from Figures \ref{fig:regM3} and \ref{fig:loglog-2M3} that 
the curve exhibits a power like decay, with an asymptotic limit to zero. In Figure \ref{fig:regM3}, we perform a log-log plot and regression taking $T=1$ and it yields $\tau_\sigma\propto\sigma^{-2/3}$, whereas the same analysis in Figure \ref{fig:loglog-2M3} taking $T=2$ and considering only those exit times that are in the interval $[1,2]$ yields $\tau_\sigma \propto\sigma^{-1}$.

{\begin{figure*}[!htbp]
   
        \includegraphics[
height=0.35\textwidth,
width=0.6\textwidth
]%
{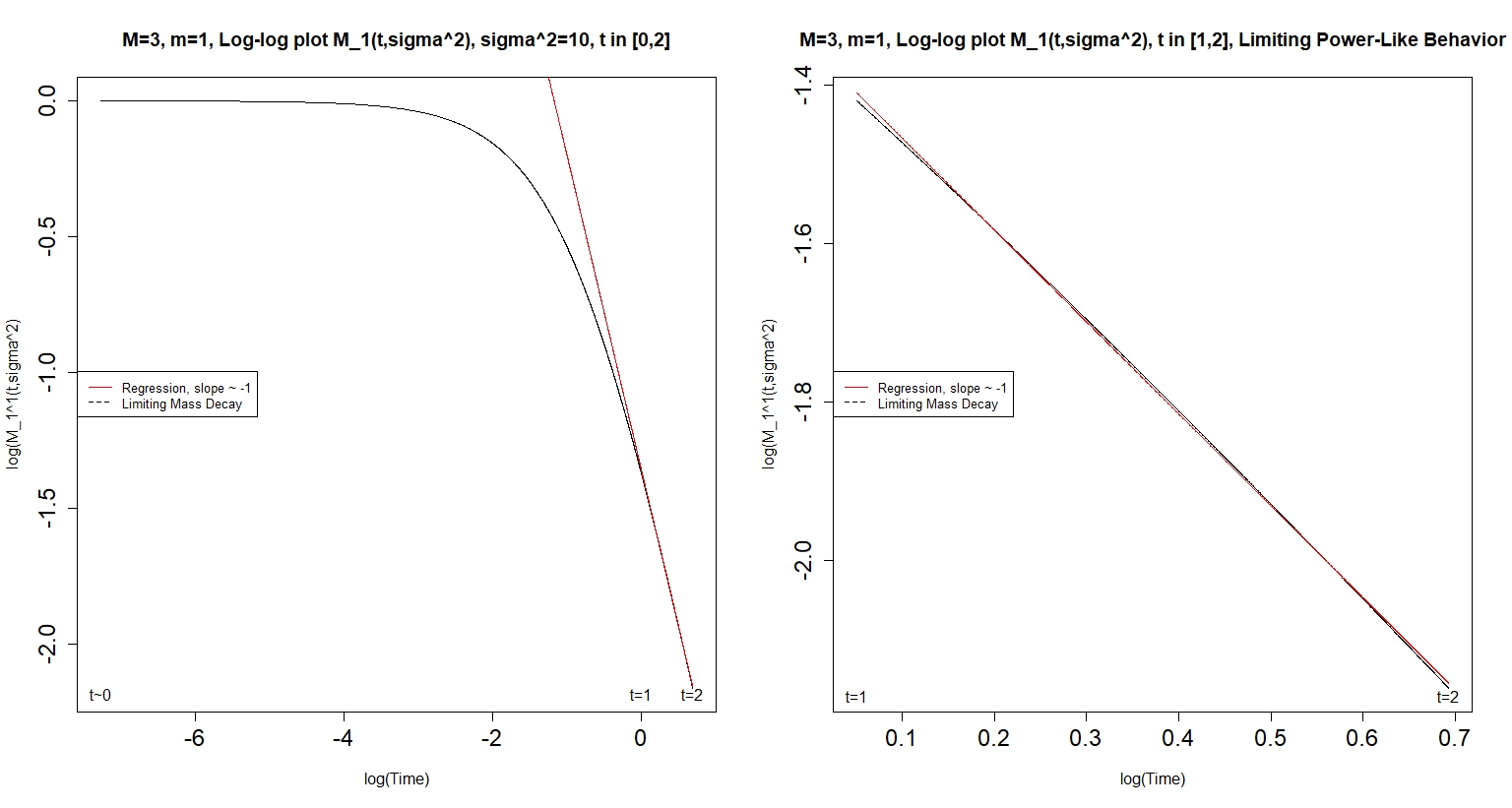}
         \caption{$M=3$, $m=1$; On the left, a plot of $\log\left(\cM_1^\sigma(t)|_{m=1}\right)$ versus $\log (t)$ in the time window $[0,2]$ at fixed $\sigma^2=10$, and on the right a close-up in the time window $[1,2]$, suggest that $t\mapsto \cM_1^\sigma(t)|_{m=1}$ is of inverse power $1$. This is consistent with the case $M=1$.}
         \label{fig:loglogM3m1}

     \end{figure*}}

{\begin{figure*}[!htbp]
   
        \includegraphics[
height=0.35\textwidth,
width=0.6\textwidth
]%
{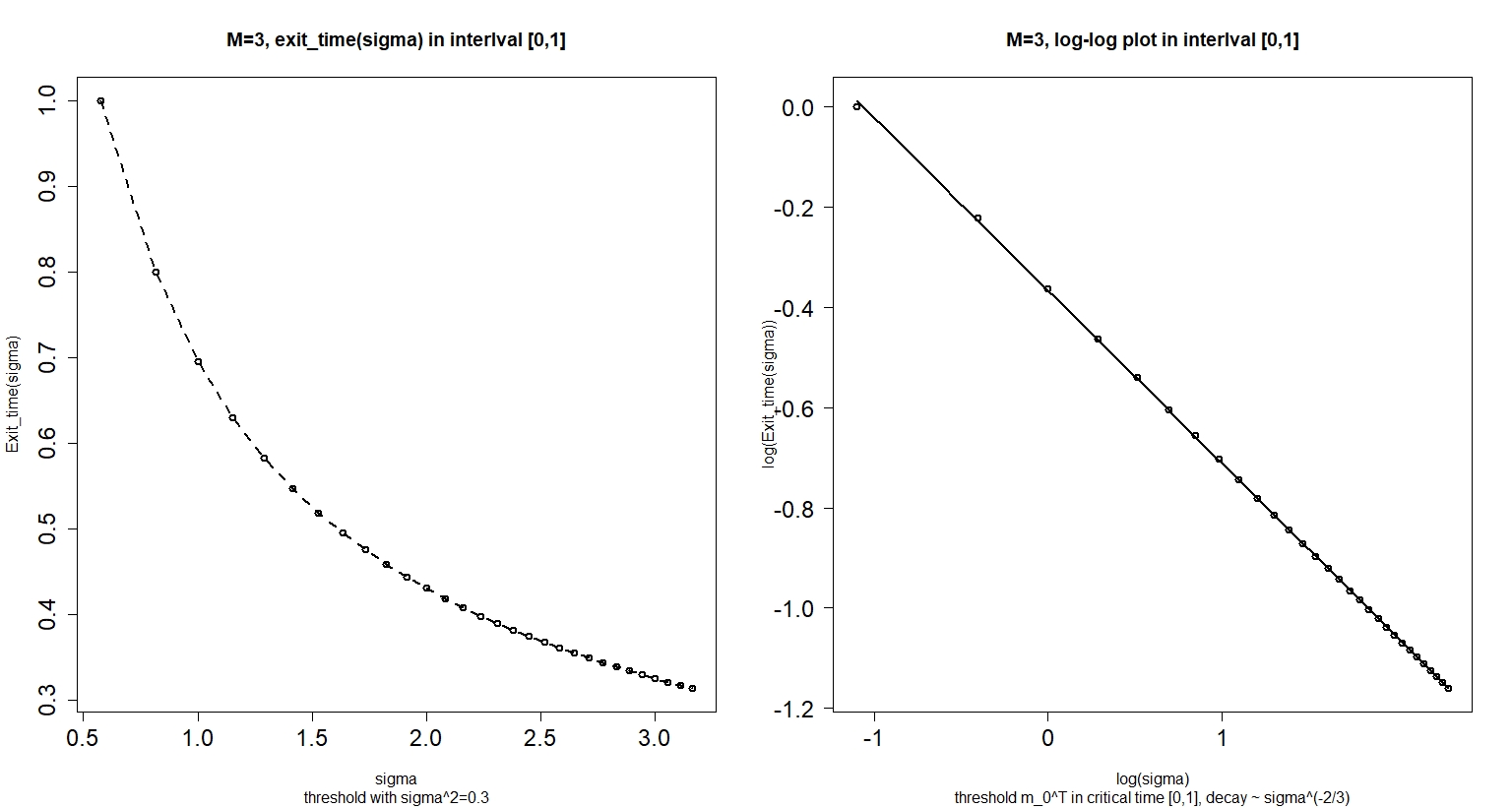}
         \caption{$M=3$; A plot of the barrier exit time $\tau_\sigma$ with respect to the turbulence parameter $\sigma$, and the corresponding log-log regression in the time window $[0,1]$ yields $\tau_\sigma \propto\sigma^{-2/3}$.}
         \label{fig:regM3}

     \end{figure*}}
     
{\begin{figure*}[!htbp]
   
        \includegraphics[
height=0.35\textwidth,
width=0.6\textwidth
]%
{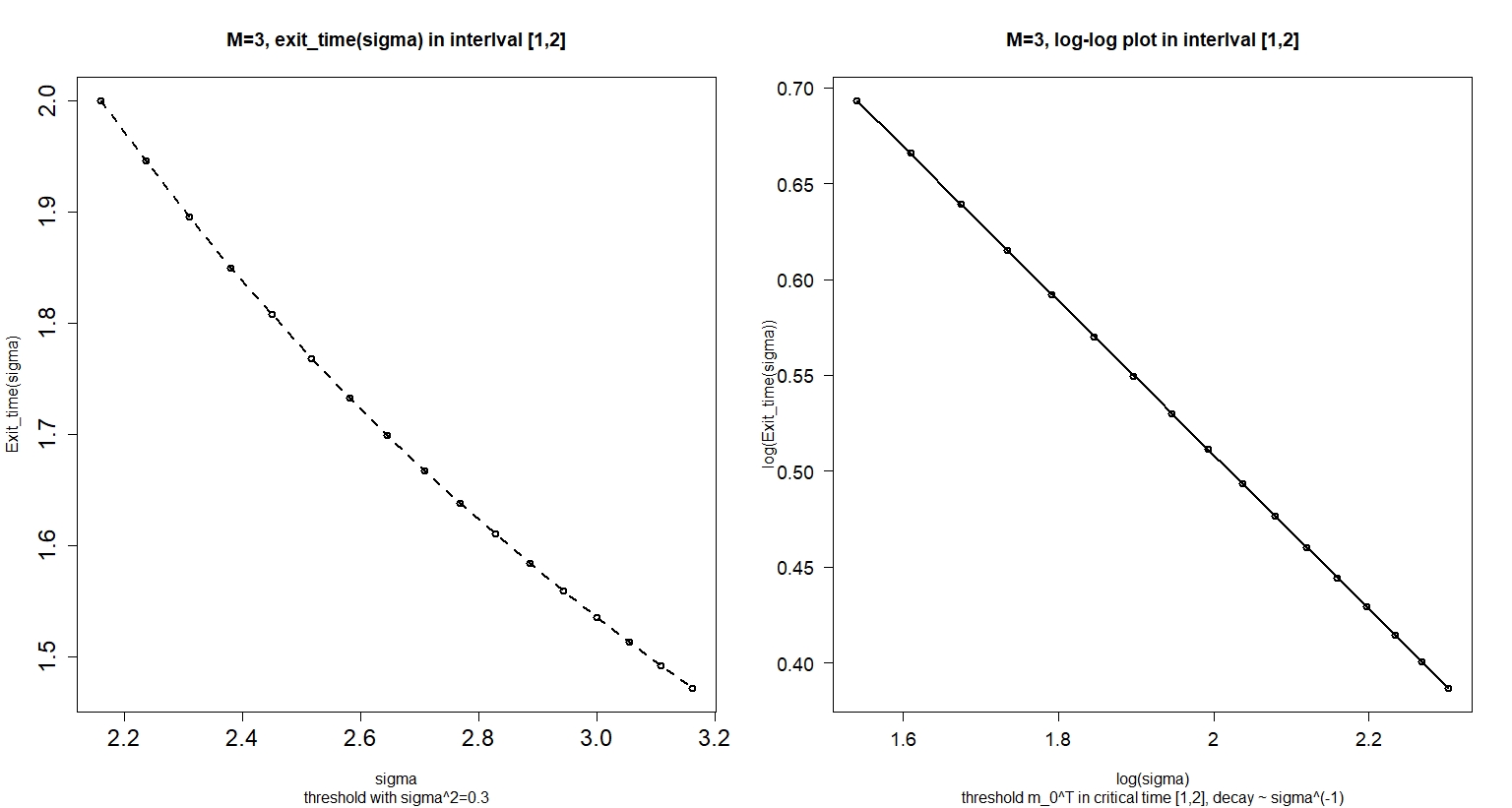}
         \caption{$M=3$; A plot of the barrier exit time $\tau_\sigma$ with respect to the turbulence parameter $\sigma$, and the corresponding log-log regression in the time window $[0,2]$, taking into consideration only those exit times in the interval $[1,2]$, yields $\tau_\sigma \propto\sigma^{-1}$.}
         \label{fig:loglog-2M3}

\end{figure*}}     
Thus, when $M>1$, and the initial mass is located on a single level, we lose the conjectured behavior of Subsection \ref{sec:M1}, and we can only expect that the function \eqref{totmass} has the same asymptotic limit as
\begin{align*}
\cM_1^\sigma(t)\gtrsim \frac{1}{A_d(\sigma) t +{\cM^\sigma_1(0)}^{-1}},
\end{align*}
for some function $A_d$ that depends on dimension $d$, and that $A_1(\sigma)\propto\sigma$. A rough explanation of this numerical finding may be the following one: when $M>1$ and the density $f\left(
0,\mathbf{v}\right)$ is in the unique level $m=1$, from (\ref{eq:collision}) we see that the poisitive part $\cQ_m^+$ is greater than the negative part $\cQ^-_m$ for a transient period of time in which, for $m>1$ mass should increase before decay, suggesting a delay, and as such a reported slower decay, of the ``total mass". Also supporting this idea are the numerical simulations performed on the rapidity of decay for level $m=1$. Here $\cQ^+_1=0$, and we see the same behavior as the limiting case in which only one type of mass is considered.

{\begin{figure}[h!]
\includegraphics[
height=0.45\textwidth,
width=0.45\textwidth
]%
{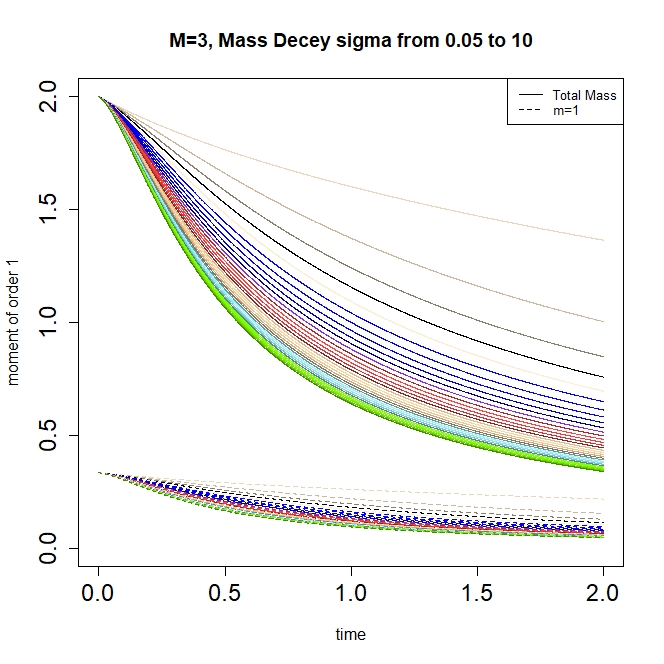}
\caption{$M=3$; Decay of $\mathcal{M}_1^\sigma(t),\ t\in[0,2]$. Initial density $f_j(0, \bv)$, $j=1,2,3$ concentrated on $\bv\in[-1/2,1/2]$, following \citep{Hammond}. The parameter $\sigma^2$ ranges in the set $0.05$ to $10$. A visible increase in coagulation is present. Dashed lines are the single mass for $m=1$, $\cM_1^\sigma(t)|_{m=1}$.}
\label{fig1M3s2}
\end{figure}
}
\subsection{Diffused mass concentration: $M>1$}
Here we propose a first analysis of the aformentioned second setting: the one that follows the theoretical assumptions as in \cite{Hammond,FHP,FlandHuang2}. In detail, the initial mass is not concentrated only in one layer, but is generated according to two probability distributions so that $\mathbb P(m_1(0)=m)=r(m)$ with $\sum_{m=1}^Mr(m)=1$, and deterministic probability densities functions $g_m(\bv)$, $m=1,2,...,M$, satisfying suitable regularity and decay assumptions, such that 
\begin{align}\label{eq:ini}
f^0_m(\bv)=r(m)g_m(\bv),\quad \forall m.
\end{align}
As such, we select initial conditions compactly supported in a small range of velocity, i.e. $[-1/2,1/2]$, to better look at the behavior of the mass decay through time. We note here that this is the natural setting that generalizes the case of $M=1$. We perform a numerical simulation of the system \eqref{finaleq} for 
dimension $d=1$, maximal mass level $M=3$ and time window $[0,2]$.

In Figure \ref{fig1M3s2}, we plot the function \eqref{totmass} for different values of the turbulence parameter $\sigma^2$ that ranges from $0.05$, that we refer to as the non-turbulent case, to $10$, which represents an intense eddy diffusivity. 
As in the case of $M=1$, it shows a faster decay correlated with the increase of turbulence, 
and a speedup coagulation process. Plotted with dotted lines we show the decay of mass $m=1$. This behavior is analogous for $m=1,2,3$.

For fixed $\sigma^2=10$ we perform a log-log plot in time window $[0, 2]$ as presented in Figure \ref{fig:loglogM3s2}. It shows that $t\mapsto \cM_1^\sigma(t)$ is of inverse power approximatly $1$, after a transient time period dependent on the finiteness of the initial condition. As conjectured in the case $M=1$, we see a consistency in the behavior of the ``total mass" when $M$ increase: the initial condition is active everywhere, maintaining the structure of a probability density, thus making the results not unexpected. In fact, analyzing the coagulation operator (\ref{eq:collision}), we see that $\mathcal{Q}^+$ is not dominant when all the masses of the system are spread over all the analyzed layers. Therefore, we see an immediate decrease in mass for $m\neq1$ and as such a maintained global decay of $\mathcal{M}_1^\sigma(t)$, the total mass.

{\begin{figure}[h!]
\includegraphics[
height=0.45\textwidth,
width=0.45\textwidth
]%
{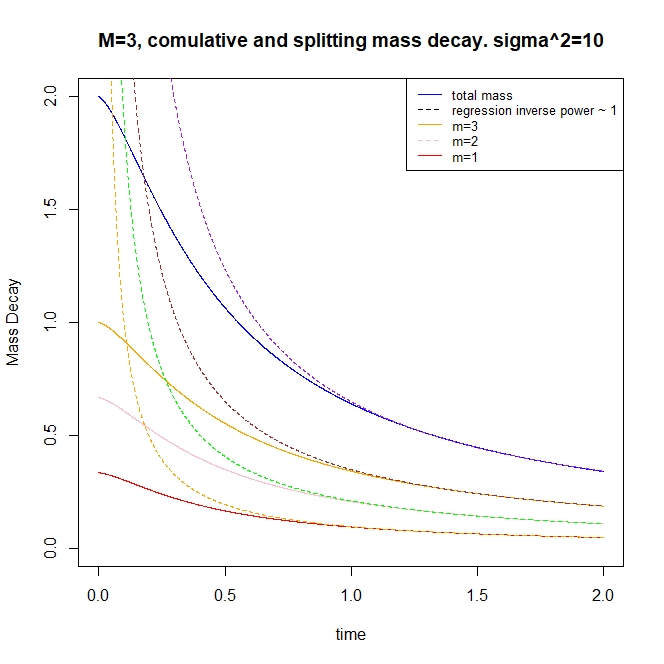}
\caption{$M=3$; $\mathcal{M}_1^\sigma(t)$ for the total mass and the single levels $\cM_1^\sigma(t)|_{m=k}$, $k\in\{1,2,3\}$ for $\sigma^2=10$. In dashed lines we see the expected limiting behaviors and the relative power of order $\approx1$, suggesting consistent log-logistic behaviors as conjectured for  system with $M<\infty$.}
\label{fig:inverse-power-1-M3s2}
\end{figure}
}
{\begin{figure*}[t!]
\includegraphics[
height=0.35\textwidth,
width=0.6\textwidth
]%
{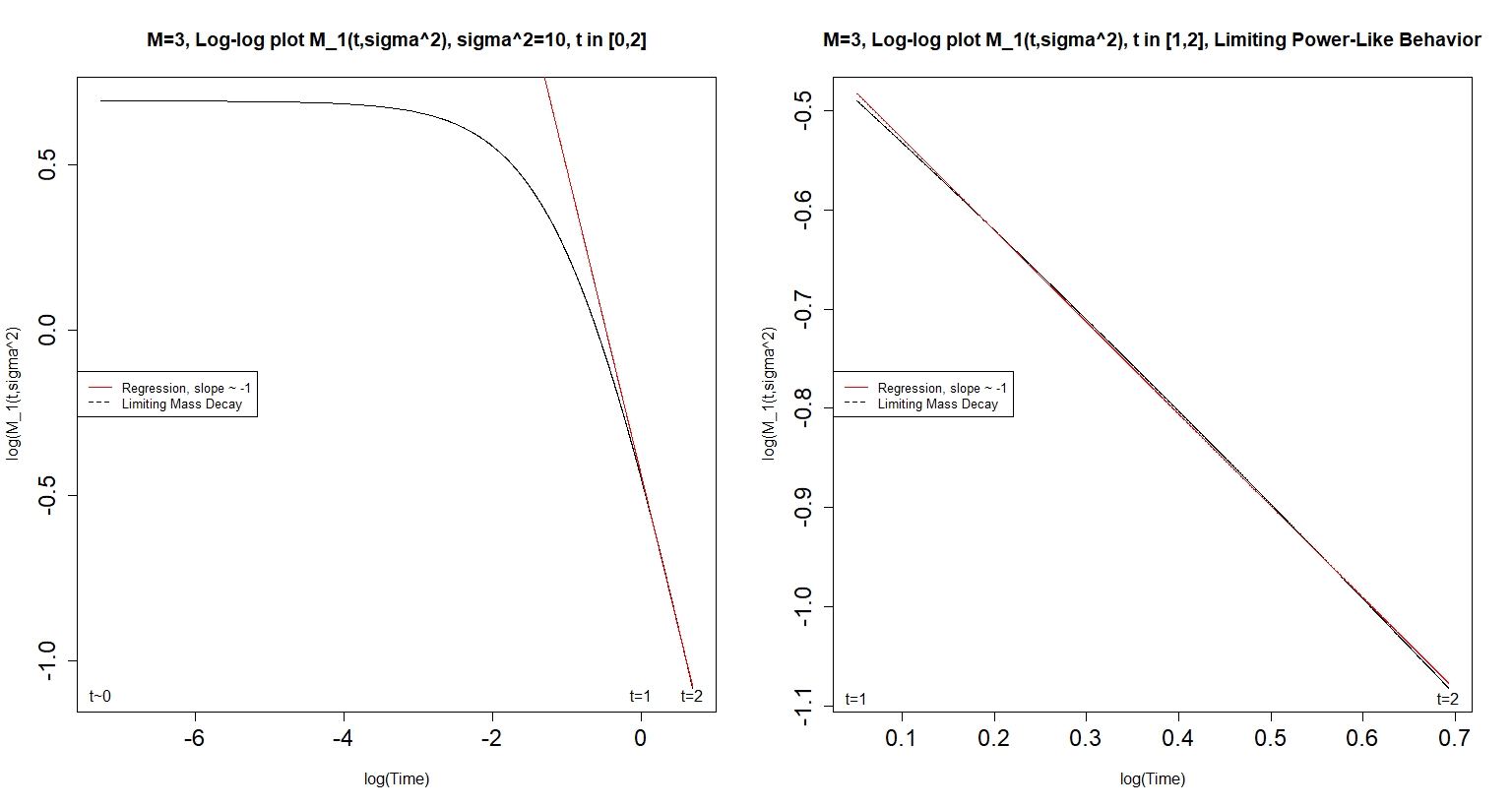}
\caption{$M=3$; On the left, a plot of $\log\left(\cM_1^\sigma(t)\right)$ versus $\log (t)$ in the time window $[0,2]$ at fixed $\sigma^2=10$, and on the right a close-up in the time window $[1,2]$, suggest that $t\mapsto \cM_1^\sigma(t)$ is of inverse power $\approx1$. A transient behavior is present due to the finite initial condition.}
\label{fig:loglogM3s2}
\end{figure*}
}
{\begin{figure*}[t!]
   
        \includegraphics[
height=0.35\textwidth,
width=0.6\textwidth
]%
{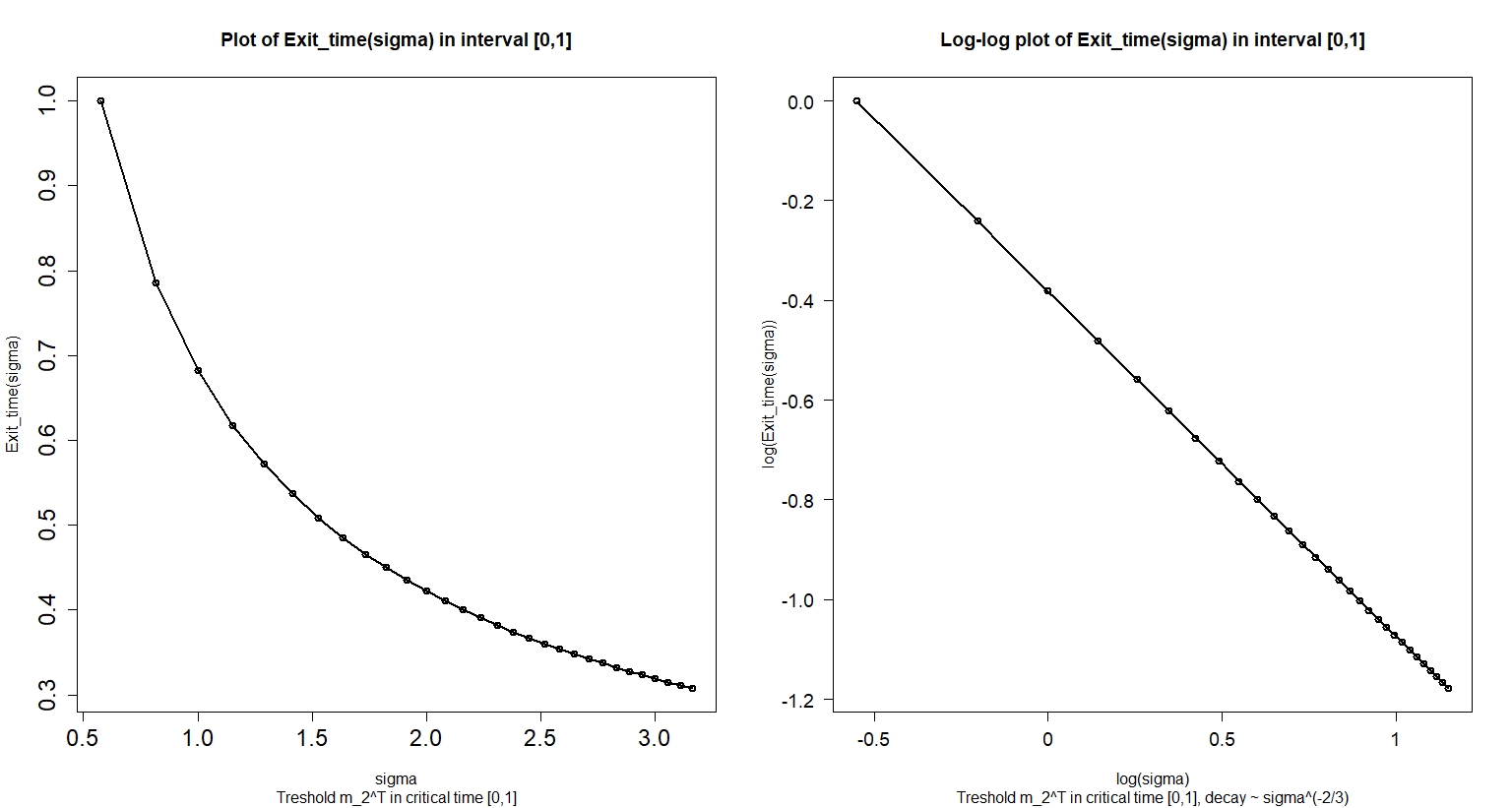}
         \caption{$M=3$; A plot of the barrier exit time $\tau_\sigma$ with respect to the turbulence parameter $\sigma$, and the corresponding log-log regression in the time window $[0,1]$ yields $\tau_\sigma \propto\sigma^{-2/3}$.}
         \label{fig:loglog-1M3s2}

\end{figure*}} 
  
For this reason, as shown in Figure \ref{fig:inverse-power-1-M3s2}, we study the decay of the single mass $m\in\{1,2,3\}$. The figure shows the regression curves plotted with dashed lines. As in Figure \ref{fig:inverse-power-1}, we maintain a relation of inverse power in time, approximately of $1$, after a transient time period.  As a further exploration, we see that the same behavior is present, 
and the curve exhibits a power like decay, with an asymptotic limit to zero.
  
Concerning the behavior of the barrier exit time, we see from Figures \ref{fig:loglog-1M3s2} and \ref{fig:loglog-2M3s2} that 
the curve exhibits a power like decay, with an asymptotic limit to zero. In Figure \ref{fig:loglog-1M3s2}, we perform a log-log plot and regression taking $T=1$ and it yields $\tau_\sigma\propto\sigma^{-2/3}$, whereas the same analysis in Figure \ref{fig:loglog-2M3s2} taking $T=2$ and considering only those exit times that are in the interval $[1,2]$ yields $\tau_\sigma \propto\sigma^{-1}$. Thus, when $M>1$, and the initial mass is spread over all the mass levels, we are close to the conjectured behavior of previous section, and we can expect that the function \eqref{totmass} has the same asymptotic limit as
\begin{align}
\cM_1^\sigma(t)\sim \frac{1}{A_d(\sigma) t +{\cM^\sigma_1(0)}^{-1}},  \label{conj:totmass}\\
\cM_1^\sigma(t)|_{m=1}\sim \frac{1}{A^1_d(\sigma) t +{\cM^\sigma_1(0)}|_{m=1}^{-1}},\label{conj:singlemass}
\end{align}
{\begin{figure*}[t!]
   
        \includegraphics[
height=0.35\textwidth,
width=0.6\textwidth
]%
{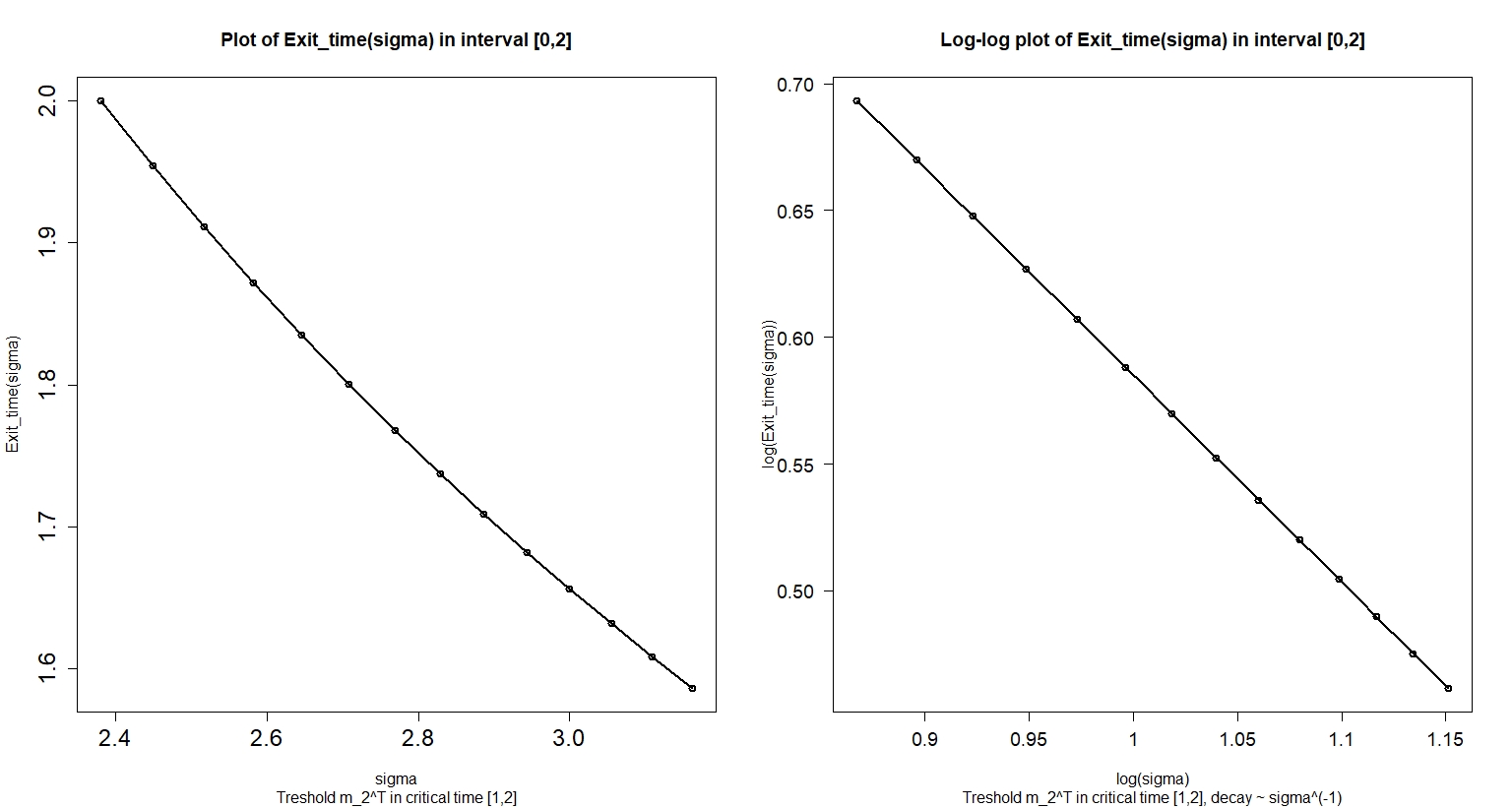}
         \caption{$M=3$; A plot of the barrier exit time $\tau_\sigma$ with respect to the turbulence parameter $\sigma$, and the corresponding log-log regression in the time window $[0,2]$, taking into consideration only those exit times in the interval $[1,2]$, yields $\tau_\sigma \propto\sigma^{-1}$.} 
         \label{fig:loglog-2M3s2}

\end{figure*}} 
for some function $A_d$ that depends on dimension $d$, and that $A_1(\sigma)\propto\sigma$. A rough explanation of this numerical finding may be the following one: when $M>1$ and the density $f\left(t,\mathbf{v}\right)$ is spread over all levels $m=1,...,M$, from (\ref{eq:collision}) we see that the positive part $\cQ_m^+$ is already negligible with respect to that of $\cQ^-_m$, for all $m$. In particular, the masses are drawn immediately 
to masses $>M$, that we interpret as falling rain outside of our system. 
Supporting this we see in Figure \ref{fig:inverse-power-1-M3s2} no transient period of time in which, for $m>1$, mass increases before decaying, suggesting no delay, and as such the decay of the ``total mass" is maintained. Note that $\cQ^+_1=0$ and, as expected, we see the same behavior as the limiting case in which only one type of mass is condidered.\\

We summarize in Table \ref{tab:params} the precise fitting obtained through non-linear regression for all the analyzed quantities. The table shows accordance with our proposed decay behavior and suggests a future analysis for different initial conditions and higher dimensions.

\begin{table}[h!]
\begin{tabular}{|c|c|c|c|}
\hline
& \textbf{$M=1$} & \textbf{$M=3$ localized} & \textbf{$M=3$ diffused}\\
\hline
\hline
\textbf{$\bm{\tau_\sigma[0,1]}$} & -0.66& -0.69& -0.68\\
\hline 
\textbf{$\bm{\tau_\sigma[1,2]}$} &-0.94& -0.92& -0.91\\
\hline
\textbf{$\bm{\mathcal{M}_1^\sigma(\sigma^2=10)}$} & -0.96 & -0.81& -0.94\\
\hline
\end{tabular}
\caption{Table showing precise fitting parameters, on a log-log scale, for the decay in time of $\bm{\mathcal{M}_1^\sigma(t)}$ and for the exit barrier $\bm{\tau_\sigma}$.}
\label{tab:params}
\end{table}

\subsection{Mean Collision Rate}
Finally, in this segment we propose numerical simulations that validate the theoretical behavior proposed in Section \ref{sec:rel-vel}. 

In particular, we have analyzed the same setting as in \ref{totmass}, which either $M=1$ or $M=3$. Computed with the procedure that we will explain below, all the case agree with equation \ref{limitDens} and the theory proposed in \ref{sec:rel-vel}. 
As such, for visual clarity, here we illustrate results in the simpler case $M=1$, with $m=1$, and compute the behavior of $R_{1,1}$ and its law respect to the fluctuation parameter of the velocity, $\sigma$. 

The same simulations, with $M=3$, focusing on different mass level $m\in\{1,2,3\}$ and different initial conditions are briefly discussed in Appendix \ref{C}, Figure \ref{AllR}. There, computed limiting value $R_{m_1,m_2}$ show accordance with the simulations with $M=1$.

From here on, we fix $\gamma=1$ since objective of the paper is the understanding of the dependence on the turbulent kinetic energy of collision rate $R_{m,m}$. However, we note that this parameter is important to the complete understanding of the behavior of this kind of systems, thanks to is relation with Stokes Number, and as such would be subject of future studies.   

We know from \ref{limitDens} that a candidate estimation for $R_{m_1,m_2}$ is obtain throughout the steady state density of the system. For this reason, concerning the simulation, independently on $M$, we selected a concentrated initial condition with moderate velocity and we let the system evolve in the time frame $t\in[0,4]$, producing solution $(f^\sigma_m(t,v))_m$.

Since no mass conservation is present for the finite system $M<\infty$, and density is moved to higher levels not preserving the starting probability, we normalize at each time step the density $f^\sigma(t,\mathbf{v})$, solution of our Smoluchowski equation, i.e. we consider
$$
\xi_1^\sigma(t,\mathbf{v}):=f_1^\sigma(t,\mathbf{v})\left(\int\ f_1^\sigma(t,\mathbf{v})dv\right)^{-1}
$$
and, with this, the product probability $\xi_1(v)dv\otimes \xi_1(w)dw$. We are able to compute a time dependent, mean in velocity, collision rate:
$$
\mathbf{R_g(t,\sigma)}:=\iint\ |\mathbf{v}-\mathbf{w}|\xi_1^\sigma(t,\mathbf{v})\xi_1^\sigma(t,\mathbf{w})dvdw
$$

\begin{figure}[!t]
    \includegraphics[width=0.45\textwidth]{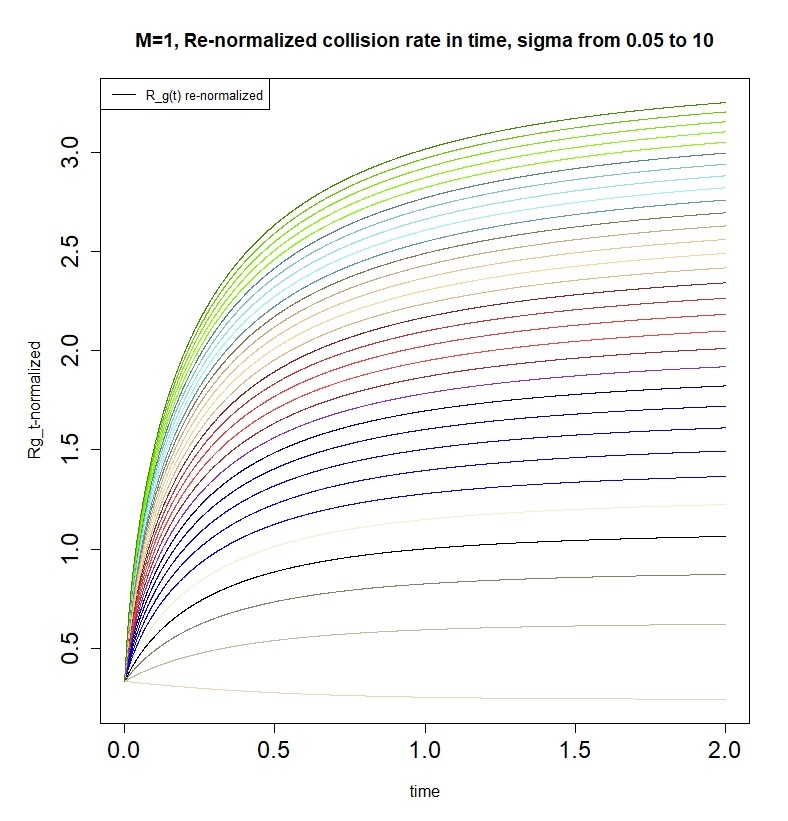}
  \caption{M=1; concentrated initial ccondition $f_0(v)$. 
  The \textbf{re-normalized time dependent collision rate} $\mathbf{R_g(t,\sigma)|_{[0,2]}}$ obtained with the new probability density $\xi_1^\sigma\otimes \xi_1^\sigma$.}
  \label{massrenorm}
\end{figure}

In Figure \ref{massrenorm}, is shown the result for $M=1,\ m=1$ and this re-normalize collision rate.
Each of the curves $R_g^\xi(\sigma)$ as an inverse behavior of a log-logistic function with exponent 1 in $\sigma$, suggesting a plateau in time.
A such, this time dependent probability distribution on the product space of the velocity domains as a limiting density and we can argue that
$$\mathbf{R_g(t,\sigma)}=\mathbb{E}_{m,m}[|\mathbf{v}-\mathbf{w}|]\rightarrow_{t\rightarrow\infty} R_{1,1}^M.$$

In fact, as shown in Appendix \ref{C}, Figure \ref{rga1}, the computed quantity $\xi(T,\mathbf{v})$ approximate the theoretical limiting density $p_1(\mathbf{v})\sim \mathcal{N}(0,\sigma^2)$, for this reason we initialize the evolving system with the proposed steady state condition $f_0^1(t,\mathbf{v}):=\xi(T,\mathbf{v})$, for different $\sigma$.

This means that we expect $\xi(T,v,\sigma)$ to be closer to the steady state distribution after a small time and the computed $\mathbf{R_g(t,\sigma)}$ will be $\propto R_{m_i,m_j}$. As such, we restart the system with this new initial condition. To take into account that the velocity is spread, with value greater than one, and the total density near this high value is not negligible in comparison with the concentrated initial condition that we used throughout our experiment, we enlarged the velocity domain and the time domain to produce stable results on the decay of the masses and also on the mean rate $R_g(t,\sigma)$. 

\begin{figure}[!t]
    \includegraphics[width=0.45\textwidth]{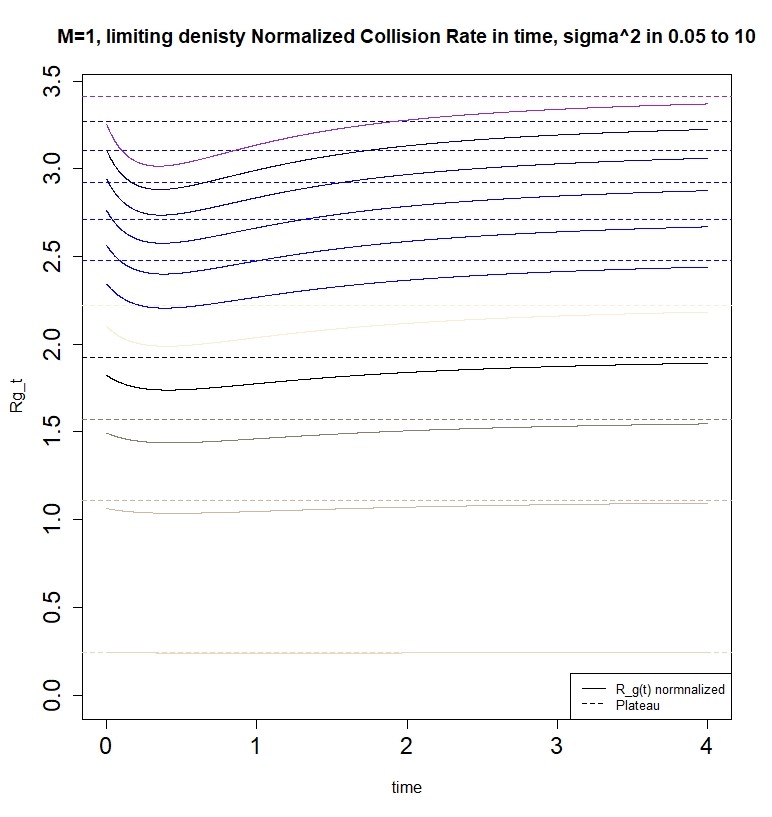}
  \caption{M=1; Initial condition $f_0(v)=\xi(T,v)$ approximation of stationary density. 
The \textbf{re-normalized time dependent collision rate} $\mathbf{R_g(t,\sigma)}$ show stationary behavior. Darker line corresponds to higher sigma in the set $[0.05,10]$.}
  \label{rg03}
\end{figure}

In Figure \ref{rg03} we show result on the re-started system, conferming the asymptotic limit of the collision rate and an increase in $\sigma$, the turbulent parameter of the system.  
We see a small flactuating period in which the rate is not increasing and than a fast stabilization that is linked to the velocity displacement of the steady state solution.
In fact the new initial density condition produce, as expected, the same decay in the mass (since this depends only on $\sigma$ and integral of the initial condition), but for a transient period the interaction kernel $Q_m(f)$ is much stronger that the speed in which diffusion of the Laplacian act, since the new initial condition is not negligible for high value of velocity. 
As such the plateau, which agrees with Figure \ref{massrenorm}, is reached after a small period of activation of the diffusion parameter.

\begin{figure}[!h]
    \includegraphics[width=0.45\textwidth]{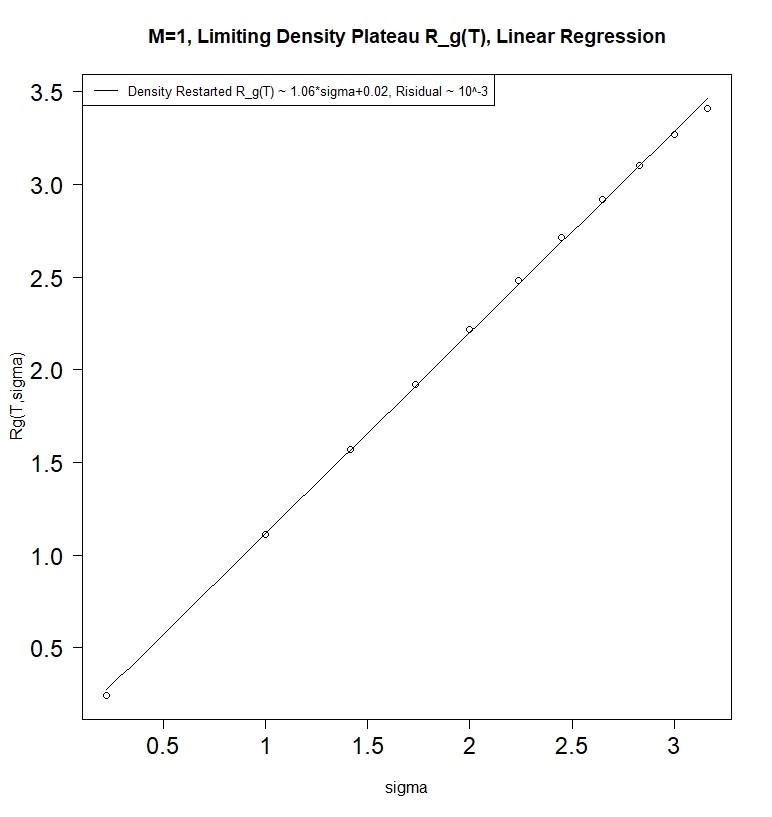}
  \caption{M=1; Initial condition $f_0(v)=\xi(T,v)$ re-normalized ending point of the simulation. Plotted limit in time $R_g(t,\sigma)$ show increase with $\sigma$. A linear regression in $\sigma$ is performed with mean error $0.001$.}
  \label{rp3}
\end{figure}

Concluding, in Figure \ref{rp3} we see that a linear relation with $\sigma$ is present with angular coefficinet near $1$, validating the expected behavior of $R_{1,1}$ with theoretical equation \ref{limitDens}. 
This is expected and in line with the previuos reasoning and also with the small transient initialization.

\section{Conclusion}
In this article, we 
presented a new kinetic model of a modified Smoluchowski \abbr{PDE} system with discrete and finite mass levels, that takes advantage of small scale turbulence and eddy diffusion in the velocity variable to enhance coagulation. We presented the derivation of the \abbr{PDE} system from a particle-fluid model subjected to a transport-type noise, and we 
analyzed numerically the behavior of its solutions.
We showed that coagulation efficiency increases steadily with the increase of turbulence and, moreover, 
a power-law decay in time and in the turbulence parameter is present.
Conluding, we have presented analytic and numerical presentation to understand the key factor of the collision rate as the average relative velocity between particles.

\begin{appendix}
\section{Derivation of \eqref{SCeq} from particle-fluid interaction}\label{appen-particle}
We present the sketch of the scaling limit to an \abbr{SPDE} from particle-fluid interaction for the truncated model (at threshold $M$). 

For any $d\ge1$ and $N, M\in\mathbb{N}$, consider an interacting particle system with space variable $\bx_i^N(t)$ in $\mathbb{T}^d$, velocity variable $\bv_i^N(t)$ in $\R^d$, mass variable $m_i^N(t)$ in a finite set $\{1,...,M\}$, and initial cardinality $N(0)=N$. Between coagulation events, the motion of an individual active particle obeys (recall \eqref{particle-fluid})
\begin{align}\label{particle-sys}
&\begin{cases}
d\bx^N_i(t)=\bv_i^N(t)dt, \\[5pt]
d\bv^N_i(t)\\
=\displaystyle{\frac{\alpha} {(m_i^N(t))^{1-1/d}}}\left[\sum_{k\in K}\sigma_k(\bx_i^N(t))\,\circ\,dW_t^k-\bv_i^N(t)dt\right]
\end{cases} \nonumber\\
& \qquad , i\in\mathcal N(t),
\end{align}
where 
\begin{itemize}
\item
$\sigma_k(\bx):\mathbb T^d\to\R^d$, $k\in K$ is a given (at most countably infinite) collection of smooth, deterministic, divergence-free vector fields.
\item
$\{W_t^k\}_{k\in K}$ is a given  collection of standard Brownian motions in $\R$.
\item
$\circ$ denotes Stratonovich integration, according to Wong-Zakai principle \cite{WoZa}.
\item
$\alpha$ is a positive constant that appears in Stokes' law, that includes the dynamic viscosity coefficient of the fluid.
\item
$\mathcal N(t)\subset\{1,2,...,N\}$ is the set of indices of particles that are still active at time $t$, with $\mathcal N(0)=\{1,2,..,N\}$.
\end{itemize}
After each coagulation, the index set $\cN(t)$ will change (decrease), and the velocity of a still-active particle $i$ will be reset according to the conservation of momentum, to be described a few paragraphs below.

We note again that the velocity component of the dynamics \eqref{particle-sys} obeys  Stokes' law for the frictional force exerted on a spherical particle immersed in a fluid, cf. \cite{Mehlig, wilkinson2006caustic}, with the fluid velocity idealized by the white noise velocity field $\mathbf U(t,\bx)$ \eqref{white-noise} (that acts simultaneously on all particles). This goes in the spririt of Kraichnan's model \cite{kraichnan1968small, kazantsev1968enhancement}.

We denote the $d\times d$ spatial covariance matrix of $\mathbf U(t,\bx)$ by 
\begin{align*}
C (\bx,\by):=\sum_{k\in K}\sigma_k(\bx)\otimes\sigma_k(\by).
\end{align*}
Moreover, for any fixed $\bx\in\mathbb{T}^d$ we denote the second-order divergence form elliptic operator, acting on suitable functions on $\R^d$
\begin{align*}
(\mathcal{L}^{C ,\bx}_v f)(\bv):=\frac{1}{2}\text{div}_v\left(C (\bx,\bx)\nabla_v f(\bv)\right).
\end{align*}
{{With suitable choice of $\{\sigma_k\}_{k\in K}$, see \cite{Galeati, FlaGaleLuoJEE, flandoli2021delayed}, we can have that 
\begin{align}\label{enhan-diss}
\mathcal{L}^{C ,\bx}_v\equiv \frac{\sigma^2}{2} \Delta_v, \quad \forall \bx.
\end{align}
}}

Each particle $i\in\mathcal N(t)$ has a mass $m_i^N(t)\in\{1,2,..,M\}$ which changes over time according a stochastic coagulation rule to be described below. The initial mass $m_i(0)$, $i=1,...,N$, are chosen i.i.d. (independent and identically distributed) from $\{1,2,..,M\}$ according to a probability distribution so that $\mathbb P(m_1(0)=m)=r(m)$ with $\sum_{m=1}^Mr(m)=1$. We are also given deterministic probability density functions $g_m(\bx,\bv):\mathbb T^d\times\R^{d}\to\R_+$, $m=1,2,...,M$, satisfying suitable regularity and decay assumptions, such that if $m_i(0)=m$ then the initial distribution of $(\bx_i(0),\bv_i(0))$ is chosen with probability density $g_m(\bx,\bv)$, independently across $i$. We denote
\begin{align}\label{eq:ini}
f^0_m(\bx,\bv):=r(m)g_m(\bx,\bv), \quad \forall m.
\end{align}

The rule of coagulation between pairs of particles is as follows. Let $\theta(\bx):\R^d\to\R_+$ be a given smooth symmetric probability density function in $\R^d$, that is, $\int\theta d\bx=1$, with compact support in $\mathbb B(0,1)$ (the unit ball around the origin in $\R^d$) and $\theta(0)=0$. Then, for any $\epsilon\in(0,1)$, denote $\theta^\epsilon(\bx):\mathbb T^d\to\R_+$ by
\begin{align*}
\theta^\epsilon(\bx):=\epsilon^{-d}\theta(\epsilon^{-1}\bx), \quad \bx\in\mathbb T^d.
\end{align*}
Suppose the current configuration of the particle system is 
\begin{align*}
\eta&=(\bx_1,\bv_1,m_1,\bx_2,\bv_2, m_2,...,\bx_N,\bv_N, m_N)\\
&\in(\mathbb T^d\cup\emptyset)^N\times(\R^d\cup\emptyset)^N\times\{1,...,M,\emptyset\}^N
\end{align*}
where $(\bx_i,\bv_i,m_i)$ denotes the position, velocity and mass of particle $i$, by convention if particle $i_0$ is no longer active in the system, we set $\bx_{i_0}=\bv_{i_0}=m_{i_0}=\emptyset$ (a cemetery state). Independently for each pair $(i,j)$ of particles, where $i\neq j$ run over the index set of active particles in $\eta$, with a rate (derived from the collision kernel as in \citep{Falkovich}, compare with \eqref{efficiency})
\begin{align}\label{rate}
s_{m_i^N,m_j^N}\frac{|\bv_i-\bv_j|}{N}\theta^\epsilon(\bx_i-\bx_j)
\end{align}
we remove $(\bx_i,\bv_i,m_i,\bx_j,\bv_j,m_j)$ from the configuration $\eta$, and then in case $m_i+m_j\le M$, we add 
\begin{align*}
\left(\bx_i,\frac{m_i\bv_i+m_j\bv_j}{m_i+m_j},m_i+m_j,\emptyset,\emptyset,\emptyset\right)
\end{align*}
with probability $\frac{m_i}{m_i+m_j}$, and instead add 
\begin{align*}
\left(\emptyset,\emptyset,\emptyset, \bx_j,\frac{m_i\bv_i+m_j\bv_j}{m_i+m_j},m_i+m_j\right)
\end{align*}
with probability $\frac{m_j}{m_i+m_j}$. We call the new configuration obtained this way by $S^1_{ij}\eta$ and $S^2_{ij}\eta$ respectively. On the other hand, in case $m_i+m_j>M$, then after removing $(\bx_i,\bv_i,m_i,\bx_j,\bv_j,m_j)$ from $\eta$ we do not add a new element.

In words, if $(i,j)$ coagulate, we decide randomly which of $\bx_i$ and $\bx_j$ is the new position of the mass-combined particle, provided that the combined mass does not exceed the threshold $M$. If the position chosen is $\bx_i$, then we consider $j$ as being eliminated (no longer active) and the new particle has index $i$; whereas if the position chosen is $\bx_j$, then we consider $i$ as being eliminated and the new particle has index $j$. On the other hand, the velocity of the mass-combined particle is obtained by the conservation of momentum as in {\it perfectly inelastic collisions}.

Note that the form of the coagulation rate \eqref{rate} is such that a pair $(i,j)$ can coagulate only if $|\bx_i-\bx_j|\le\epsilon$, that is, their spatial positions have to be $\epsilon$ -close. We are interested in the case when $\epsilon=\epsilon(N)\to0$ as $N\to\infty$, so that the interaction is not of mean-field type, but local. Correspondingly, the final equation we get (see \eqref{smol-spde}) is local in the $\bx$ variable. In particular, choosing $\epsilon=O(N^{-1/d})$ ensures that each particle typically interacts with a bounded number of others at any given time, which is the analogue in our continuum context, of nearest-neighbor or bounded-range interactions common in interacting particle systems defined on lattices, see \cite{KL} and references therein. 

The essential feature of our coagulation rate is the presence of $|\bv_i-\bv_j|$, which results in the same velocity difference appearing in the limit \abbr{PDE} \eqref{SCeq}. Although such rates are widely accepted in the physics literature on rain formations, our approach views $\bv$ as an active variable; we do not approximate it by a constant that depends on other physical parameters. Diffusion enhancement feeds back on coagulation enhancement through the presence of this velocity difference. As such, our Smoluchowski equation is new with respect to existing literature.

For each $N\in\mathbb N$, $T\in(0,\infty)$ and $m\in\{1,..,M\}$, we denote the process of empirical measure on position and velocity of mass-$m$ particles in the system by 
\begin{align}\label{emp-meas}
\mu^{N,m}_t(d\bx,d\bv):&=\frac{1}{N}\sum_{i\in\mathcal N(t)}\delta_{\left(\bx_i^N(t), \bv_i^N(t)\right)}(d\bx, d\bv)1_{\{m_i^N(t)=m\}}\nonumber\\
&\in\cM_{1,+}(\mathbb T^d\times\R^d)
\end{align}
where $\cM_{1,+}:=\cM_{1,+}(\mathbb T^d\times\R^d)$ denotes the space of subprobability measures on $\mathbb T^d\times\R^d$ equipped with weak topology. The choice of the initial conditions for our system implies that $\mathbb P$-a.s.
\begin{align*}
\mu_0^{N, m}(d\bx,d\bv)\Rightarrow f^0_m(\bx,\bv)dxdv, \quad \text{as }N\to\infty
\end{align*}
for $m=1,...,M$, where $\Rightarrow$ indicates weak convergence of probability measures, and the limit $f^0_m$ \eqref{eq:ini} is absolutely continuous. We conjecture that, under the assumption of local interaction, i.e.
\begin{align}\label{local-int}
\lim_{N\to\infty}\epsilon(N)=0, \quad \limsup_{N\to\infty}\frac{\epsilon(N)^{-d}}{N}<\infty,
\end{align}
for every finite $T$, the collection of empirical measures $\left\{\mu^N_t(d\bx,d\bv):t\in[0,T]\right\}_{m=1}^M$ converges in probability, as $N\to\infty$, in $\mathcal D\left([0,T],\cM_{1,+}\right)^{\otimes M}$, where $\mathcal D\left([0,T],\cM_{1,+}\right)$ is the space of c\`adl\`ag functions taking values in $\cM_{1,+}$ equipped with the Skorohod topology, towards an absolutely continuous limit $\left\{f_m(t,\bx,\bv):t\in[0,T]\right\}_{m=1}^M$. which is the pathwise unique weak solution to a Smoluchowski-type \abbr{SPDE} system \eqref{smol-spde}. The latter \abbr{SPDE} degenerates to the \abbr{PDE} system we study in this paper \eqref{SCeq} when the It\^o term is switched off.  Through recent progresses in stochastic fluid mechanics, cf. \cite{Galeati, FlaGaleLuoJEE, flandoli2021delayed, flandoli2022eddy, gess2021stabilization}, there exist specific limiting procedures that allow, in principle, to obtain the \abbr{PDE} from the \abbr{SPDE} by carefully choosing the vector fields $\{\sigma_k(\bx)\}_{k\in K}$. While we do not provide a rigorous proof here,
we think that this heuristic argument is sufficient to justify our interest in studying our \abbr{PDE} system.

\begin{widetext}
\begin{align}
\begin{cases}\label{smol-spde}
df_m(t,\bx,\bv)&=\left(-\bv\cdot\nabla_x+\gamma_m\text{div}_v\left(\bv\cdot\right)+\displaystyle{\frac{\gamma^2_m\sigma^2}{2}}\Delta_v\right)f_m(t,\bx,\bv)dt \\[5pt]
&\quad -\gamma_m\sum_{k\in K}\sigma_k(\bx)\cdot\nabla_v f_m(t,\bx,\bv)\,dW_t^k+\left(\mathcal Q_m^+-\mathcal Q_m^-\right)(\mathbf f, \mathbf f)(t,\bx,\bv). \\[10pt]
f_m(\cdot,\bx,\bv)|_{t=0}&=f^0_m(\bx, \bv), \quad m=1,...,M.
\end{cases}
\end{align}
\end{widetext}

\section{Explanation of the link \eqref{def sigma}}\label{appen:link}
Recall the stochastic equation \ref{spde-krai}. In real turbulent fluids, the fluid vector field $\bU(t)$ is not exactly white in time, but has a correlation length approximately $\tau_\bU$. Alleviating notations, let us only analyze the transport term involving $\bU(t)$ and introduce a time delay of duration $\tau_\bU$:
\begin{align}\label{huris}
&\gamma_m\operatorname{div}_{v}\left(  \left(
\mathbf{U}(t) \right)  f_{m}(t) \right)=\gamma_m 
\mathbf{U}(t)\nabla_vf_{m}(t) \nonumber\\
&=\gamma_m 
\mathbf{U}(t)\nabla_vf_{m}(t-\tau_\bU)+\gamma_m 
\mathbf{U}(t)\nabla_v\left(f_{m}(t)-f_m(t-\tau_\bU)\right) \nonumber\\
&=\gamma_m 
\mathbf{U}(t)\nabla_vf_{m}(t-\tau_\bU)\nonumber\\
&\quad -\gamma_m 
\mathbf{U}(t)\nabla_v\left(\int_{t-\tau_\bU}^t\gamma_m\bU(s)\nabla_vf_m(s)ds\right)+\text{other terms},
\end{align}
where the first equality is due to $\bU(t)$ independent of $v$, and 
in the last line we applied the equation \ref{spde-krai} a second time (assuming the other terms are minor).

In the limit $\tau_\bU\to0$, $\bU(t)$ approaches white noise in time, the first term of \eqref{huris} yields a local-martingale, the It\^o term. From the second term of \eqref{huris} emerges a second-order elliptic operator
\[
-\gamma_m^2\nabla_v\left(\int_{t-\tau_\bU}^t\bU(t)\otimes\bU(s)\nabla_vf_m(s)ds\right)
\]
that in the limit $\tau_\bU\to0$ is expected to converge to 
\[
-\frac{1}{2}\gamma_m^2\text{div}_v\left(C(\mathbf {0})\nabla_vf_m(t)\right)=-\frac{1}{2}\gamma_m^2\sigma^2\Delta_v f_m(t).
\]
Since the turbulence kinetic energy $k_T$ is the half-trace of the
velocity covariance tensor \cite{dupuy2019effect}, idealizing the tensor structure of $\bU(t)\otimes\bU(s)$ with $|t-s|\le \tau_\bU$, we may have that 
\[
\frac{1}{2}\bU(t)\otimes\bU(s) \sim \frac{k_T}{d}I_d, \quad |t-s|\le \tau_\bU
\]
and consequently,\[
\frac{1}{2}\int_{t-\tau_\bU}^t\bU(t)\otimes\bU(s) ds\sim \tau_\bU\frac{k_T}{d}I_d.
\]
This yields  $\frac{\sigma^2}{2}=\frac{2\tau_\bU k_T}{d}$ as claimed in \eqref{def sigma}. 

In the above argument, it is crucial that we can take $\tau_\bU$ very small while having $\gamma$ of order $1$. With $St=1/(\gamma \tau_\bU)$, the argument thus works only when $St$ is very large, and the regime where $St$ is of order $1$ requires a different analysis, consistent with the findings of \cite{abrahamson1975collision, Falkovich, Mehlig, wilkinson2006caustic}.

\section{Mean Collision Rate $M>1$ and Guassianity assumption}\label{C}
\begin{figure}[!h]
    \includegraphics[width=0.45\textwidth]{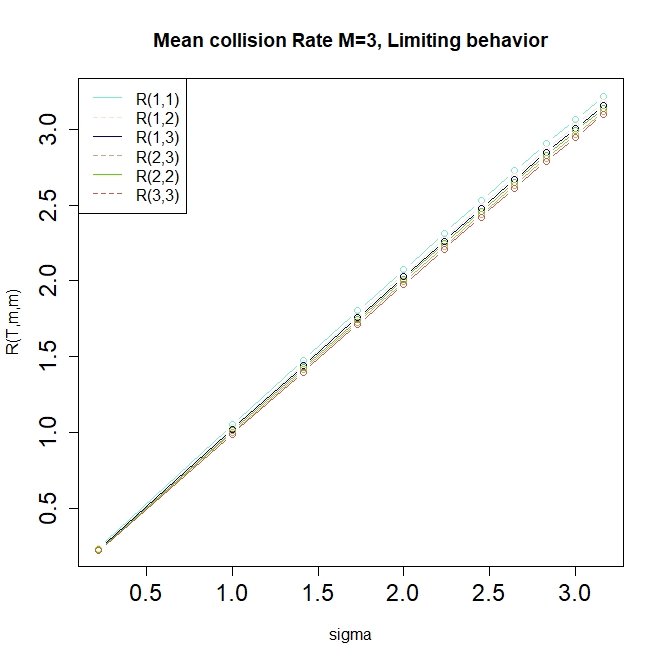}
  \caption{M=3; Plotted estimated $R_{m_1,m_2}(\sigma)$ with $m_j\in\{1,2,3\}$. A linear dependence in $\sigma$ is performed with mean error between $10^{-2}$ and $10^{-3}$.}
  \label{AllR}
\end{figure}
\begin{figure*}[!t]
    \includegraphics[width=\textwidth]{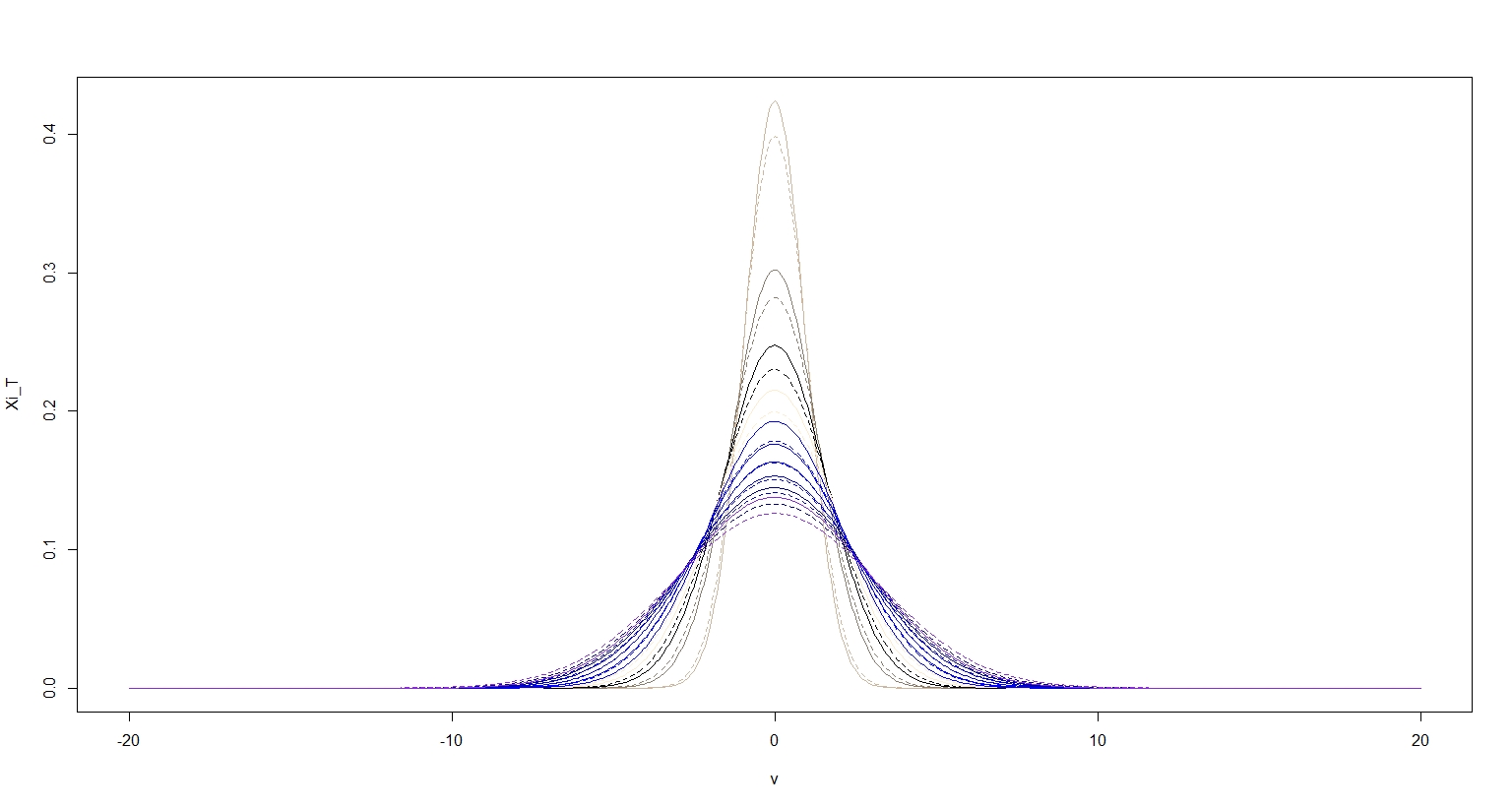}
  \caption{Solid line $\xi(T,\sigma)$ where darker colors means higher $\sigma$. Dashed lines are the Gaussian densities $\mathcal{N}(0,\sigma^2)$. The supremum norm and the $L^2$ norm of the difference differ from zero around $5\%$ to $10\%$ respectivly.}
  \label{rga1}
\end{figure*}

Using the same method proposed in Section $\ref{sec:numerical}$, we obtain analogous result for $M=3$. We analyzed two initial condition: a localized one in the mass $m=1$, and a theoretical one following \cite{Hammond}. In both of this case we used the restarting limiting density $\xi_T$ either averaged $\{\frac{1}{M}\xi_T^1,\frac{1}{M}\xi_T^2,\frac{1}{M}\xi_T^3\}$ or localized $\{\xi_T^1,0,0\}$ obtaining analogous results. In Figure \ref{AllR}, the case of localized density is shown with all combination of Collision Rate, showing agreement with the theory.

Finally, in Figure \ref{rga1}, we show the comparison between expected steady state probability and computed starting stationary solution $\xi_T(v)$, in the case $M=1$ and $T=4$. The differecne in $L^2$ norm of the two functio is less then $10^{-1}$, as per the difference between thoeretical $R_{m_i,m_j}$ and computed $R_{m_i,m_j}(T)$ estimated in less than $10^{-2}$, showing the same linear behavior.

\end{appendix}

\bibliography{biblioraindrop2022}

\end{document}